\documentclass[12pt]{article} \usepackage{latexsym} \textwidth 15cm
\begin{document}

\title{{\bf NO ASTROPHYSICAL DYADOSPHERES}
\thanks{Alberta-Thy-12-05, astro-ph/0605432}}
\author{
Don N. Page
\thanks{Internet address:
don@phys.ualberta.ca}
\\
Institute for Theoretical Physics\\
Department of Physics, University of Alberta\\
Edmonton, Alberta, Canada T6G 2J1\\
and \\
Asia Pacific Center for Theoretical Physics (APCTP)\\
Hogil Kim Memorial Building \#519\\
POSTECH, San 31\\
Hyoja-dong, Namgu, Pohang\\
Gyeongbuk 790-784, Korea
}
\date{(2006 May 18)}
\maketitle
\large

\begin{abstract}

\baselineskip 18 pt

It is shown how pair production itself would prevent the astrophysical
formation of dyadospheres, hypothetical regions where the electric field
exceeds the critical value for rapid Schwinger pair production.  Pair
production is a self-regulating process that would discharge a growing electric
field, in the example of a hypothetical collapsing charged stellar core, before
it reached 6\% of the minimum dyadosphere value, keeping the pair production
rate more than 26 orders of magnitude below the dyadosphere value, and keeping
the efficiency below $2\times 10^{-4}\sqrt{M/M_\odot}$.

\end{abstract}
\normalsize

\baselineskip 18 pt
\newpage

\section{Introduction}

Ruffini and his collaborators
\cite{DR,Ruf98,PRX,RSWX,RX,Ruf,BRX,RBFCX,RBFXC,RV,RXBFC,RBCFX,CMRX,RBBCFX,RVX,
RBCFVX,RBCFGX,RFVX,BR,RBBVXCFG,BBCFRX,FBBCRX,CBBCFRX,RBBCFGVX,RBBCFGLVX,XRFV,
RBBCFGX,Vag} have proposed a model for explaining gamma ray bursts that
presumes the initial existence of what they call a {\it dyadosphere}, a
macroscopic region of spacetime where the electric field exceeds the critical
electric field $E_c \equiv m_e^2 c^3/(\hbar e) \approx 1.32 \times 10^{16}
\:\mathrm{V/cm}$ for sufficiently rapid Schwinger pair production \cite{Sch}. 
The difficulty of producing these large electric fields is a problem with this
model that has not been adequately addressed.

There are at least two strong reasons for doubting that such large electric
fields can develop over astrophysical scales (i.e., over length scales much
larger than the collision regions of individual charged particles).  First, it
would be very difficult to develop sufficient charge imbalance for macroscopic
electric fields to produce significant numbers of pairs.  Second, even if
macroscopic pair production could somehow be achieved, I shall show in this
paper that this process is sufficiently self-regulating that it prevents the
electric field from achieving a value that would produce pairs at even
$10^{-26}$ that of dyadosphere models.  I conclude that it is highly
implausible that dyadospheres can form in outer space, and therefore invoking
them for models of gamma ray bursts is not viable.

The first reason for being extremely doubtful of the existence of astrophysical
dyadospheres is that it is very difficult for a large charge imbalance to
develop astrophysically, because of the very high charge-to-mass ratio of
elementary particles.  For example, the ratio of the electrostatic repulsion to
the gravitational attraction of two protons, each of charge $q$ and mass $m_p$,
is the square of their charge-to-mass ratio in Planck units, which is
 \begin{equation}
 \left({q\over m_p}\right)^2 \equiv {q^2\over 4\pi\epsilon_0 G m_p^2}
 \approx 1.24\times 10^{36}.
 \label{eq:a}
 \end{equation}
This implies that if one had a spherical object, such as a stellar core, with a
positive charge-to-mass ratio $Q/M$ greater than the inverse of the
charge-to-mass ratio of the proton, $m_p/q \approx 9\times 10^{-19}$, the
electrostatic repulsive force on the protons at the surface would be greater
than the gravitational attractive force, so such protons would most likely be
ejected.  (If the object had a negative charge, electrons of mass $m=m_e$ would
be expelled if $-Q/M > m_e/q \approx 4.9\times 10^{-22}$, lower by the factor
of the mass ratio of the proton and the electron, $m_p/m_e \approx 1836$, so
the maximum value of the charge of such an object would be even less if it were
negative.)

If one takes the mass-to-charge ratio of the proton to be a rough estimate of
the maximum charge-to-mass ratio of an astrophysical object (or else gravity
would not be strong enough to hold in the protons that make up the excess
charge), then using the formulas of the succeeding sections, one can readily
calculate that at the surface of a spherical object of radius $R$ and mass $M$,
the ratio of the electric field value, $E$, to the critical field value of a
dyadosphere, $E_c \equiv m_e^2 c^3/(\hbar q) \approx 1.32328539 \times 10^{16}
\:\mathrm{V/cm}$, is
 \begin{equation}
 {E\over E_c} \leq {\hbar c\over 4 G M_\odot}{m_p\over m_e}
 \left({M_\odot\over M}\right)\left({2GM\over c^2 R}\right)^2
 < 1.2\times 10^{-13}\left({M_\odot\over M}\right).
 \label{eq:b}
 \end{equation}
Therefore, if protons can be ejected from an astrophysical object whenever the
electrostatic repulsion exceeds the gravitational attraction, then the electric
field is constrained to be more than 13 orders of magnitude smaller than the
critical value for a dyadosphere (if the mass is greater than 1.2 solar masses,
which would be a conservative lower limit on any mass that could contract to
$2GM/(c^2 R) \sim 1$).  For a negatively charged object, the corresponding
limit would be more than 16 orders of magnitude smaller than the critical value
for a dyadosphere.

Although it seems very unlikely to work, one might seek to evade the
electrostatic expulsion of protons by postulating that they are bound by
nuclear forces to an astrophysical object, such as a collapsing neutron star
core.  The critical electric field $E_c$ that Ruffini and his collaborators
\cite{DR,Ruf98,PRX,RSWX,RX,Ruf,BRX,RBFCX,RBFXC,RV,RXBFC,RBCFX,CMRX,RBBCFX,RVX,
RBCFVX,RBCFGX,RFVX,BR,RBBVXCFG,BBCFRX,FBBCRX,CBBCFRX,RBBCFGVX,RBBCFGLVX,XRFV,
RBBCFGX,Vag}
used to define the minimum value for a dyadosphere would give the
electrostatic force on an electron or proton of magnitude $F_c = qE_c =
(m_ec^2)^2/(\hbar c) \approx 0.00132$ MeV/fm ($2\pi$ times the rest mass energy
$m_ec^2$ of an electron divided by the Compton wavelength $2\pi \hbar c/(m_e
c^2)$ of the electron), whereas nuclear energies of the order of an MeV (e.g.,
the deuteron binding energy of about 2.225 MeV) over length scales of the order
of a fermi (e.g., the proton charge radius of about 0.87 fm) would give nuclear
forces of the order of a few MeV/fm, about 3 orders of magnitude larger than
the electrostatic force of a minimal dyadosphere on a proton.  Then one might
suppose that the gravitational force can be provided by the neutrons of a
collapsing neutron star core, and then a tiny fraction of the nucleons are in
the form of protons that are bound by nuclear forces to the surface of the
neutron star core, a sufficient fraction to give the electric field of a
dyadosphere.

My suspicion is that such a configuration would be highly unstable to pieces of
the charged surface breaking off and being ejected by the huge electrostatic
forces on them, particularly as the core collapses and cannot maintain a fixed
configuration of its nucleons (say a crystalline structure with strong nuclear
forces between all neighboring nucleons).  I would think that it is more
reasonable that any strong nuclear crystalline structure would develop cracks
that would make the effective nuclear forces too weak to prevent the charged
surface of the core from breaking off and being ejected by an electric field
anywhere near dyadosphere values.  However, I do not have a firm result of when
this would definitely happen, and therefore I was led to do the calculations
below with the second mechanism (self-regulation) for preventing the occurrence
of a dyadosphere.

As an aside, let me describe one potential mechanism for ejecting the charge if
protons are bound by nuclear forces to a collapsing core, though it proved not
to be sufficient.  This was the idea that radiation would convert the protons
to neutrons, positrons, and anti-neutrinos, and then the positrons (not held to
the core by nuclear forces) would immediately be ejected by the strong
electrostatic forces.  Presumably there must be enough radiation present to
prevent the electrons in the inevitable plasma around the core from being
pulled back to it to neutralize it.  Then the hope was that radiation strong
enough to keep away the electrons would also be strong enough to convert the
charge carriers from protons (bound to the core) to positrons (not bound). 
However, this would be a weak interaction rate, and it turned out that the
cross section \cite{Czar} is apparently not large enough for the minimal amount
of thermal radiation needed to keep the electrons at bay to be sufficient to
convert the protons to positrons within the collapse time of the core.

Therefore, I was led to do an analysis of the self-regulation of the pair
production process itself, which, as we shall see below, will discharge any
growing electric field well before it reaches dyadosphere values.  This occurs
essentially because astrophysical length scales are much greater than the
electron Compton wavelength, which is the scale at which the pair production
becomes significant at the critical electric field value for a dyadosphere. 
Therefore, the electric field will discharge astrophysically even when the pair
production rate is very low on the scale of the electron Compton wavelength.

The calculations below lead to the conclusion that it would be very difficult
astrophysically to achieve, over a macroscopic region comparable to the size of
a black hole or larger star, electric field values greater than a few percent
of the minimum value for a dyadosphere, if that.  The Schwinger pair production
itself would then never exceed $10^{-26}$ times the minimum dyadosphere value.

\section{Schwinger discharge of a charged collapsing core}

In this section we shall analyze the pair production and discharge of an
electric field produced by the collapse of a hypothetical charged sphere or
stellar core, ignoring the processes already discussed that would most probably
cause almost all of the excess charge of the sphere to be electrostatically
ejected.

For a sphere collapsing in finite time, there is only a finite time for the
electric field to be discharged by the Schwinger pair production process, so
the discharge is never complete, but instead leaves a residual electric field
at each moment of the collapse.  We shall calculate an upper limit for this
value and show that it is always more than a factor of 18 less than the minimum
value for a dyadosphere.  Because the pair-production rate is exponentially
damped by the inverse of the electric field value, the pair production rate
itself never exceeds a value that is more than 26 orders of magnitude below
that of a dyadosphere.  (The factor of the order of $10^{26}$, which we shall
derive below, comes mainly from the fine structure constant multiplied by the
square of the ratio of the Schwarzschild radius of a solar mass black hole to
the Compton wavelength of an electron, which is why this factor is so large.)

The maximum electric field is smaller for cores that collapse into larger black
holes (because the discharge time during infall is greater), so for a very
conservative upper limit on the electric field, we shall assume that the black
hole that forms has one solar mass.  Of course, we expect that the minimum mass
of a black hole that forms astrophysically has significantly more than one
solar mass, so the corresponding maximum electric field would be weaker (by a
logarithmic factor such that the maximum pair production rate would go
essentially inversely with the square of the mass of the black hole).  That is,
we are assuming that one solar mass is a very conservative lower limit on the
mass of any black hole that forms astrophysically in the present universe (as
distinct from, say, primordial black holes that might have formed much smaller
in the very early universe; our calculations will not apply to those, but they
will have by now had a very long time also to discharge and so would also not
be expected to have significant charge today).

The maximum electric field is also smaller the slower that the core collapses
(giving more time for discharge), so again to get a very conservative upper
limit on the electric field, we shall assume that the core falls in as fast as
is astrophysically possible, which is free fall with zero binding energy.  We
assume that it is not astrophysically possible to have the spherical outer
boundary of a collapsing core moving inward with a velocity so high that it
would have come from an unbound configuration with nonzero inward velocity at
radial infinity, or that at smaller radii it could have been accelerated inward
at greater than gravitational accelerations.

Thirdly, the maximum electric field is of course smaller the smaller the
initial charge on the collapsing core.  To get the maximum electric field
possible, we shall assume that the initial charge-to-mass ratio of the
collapsing core is unity, the largest value possible for which the
electrostatic repulsive forces do not overwhelm the total gravitational
attractive forces on the entire core and prevent the core from collapsing
(since we are excluding the possibility that the core is shot in from far away
or is otherwise pushed inward by nongravitational forces, which is not at all
astrophysically plausible).

Now of course if the charge-to-mass ratio of the core really were unity, the
core would not fall until there is some discharge.  But merely to get a
conservative upper limit on the electric field as the core collapses, we shall
for simplicity assume that the initial charge equals the mass but ignore the
electrostatic repulsion, so that the core nevertheless falls in at the rate it
would from purely gravitational free fall from infinity, with no reduction of
the free fall rate by the electrostatic repulsion of the charge.  The
gravitational effects of the electric field outside the core would also reduce
the free fall time, but we shall ignore this effect as well and simply take the
external gravitational field to be given by the vacuum spherically symmetric
Schwarzschild metric.

First we shall discuss the physical quantities involved, quote the formula for
the pair production rate, and cite the definition of a dyadosphere.  Next, we
derive a crude but very tiny upper limit on the maximum pair production rate of
any macroscopic process and then make a rough estimate of the discharge rate
for a collapsing charged core.  Finally, we shall derive and solve the
differential equations describing the process more precisely, in order to give
the maximum electric field and pair production rate and the efficiency of the
process.  We also confirm that the interactions between the pairs produces are
utterly negligible.

\subsection{Planck units and physical quantities}

Since the process involves both gravity (for the free-fall collapse of the core)
and quantum field theory (for the discharge by Schwinger pair production), it is
simplest to use Planck units, in which

 \begin{equation}
 G = \hbar = c = 4\pi\epsilon_0 = k_\mathrm{Boltzmann} = 1.
 \label{eq:1}
 \end{equation}

We can then always return to conventional units by using dimensional analysis to
insert the right powers of the Planck units that we are setting equal to unity:
\begin{eqnarray}
1 \!\!&=&\!\! l_P = \sqrt{\hbar G/c^3}
 \approx 1.61624\times 10^{-33} \:\mathrm{cm},
 \nonumber \\
1 \!\!&=&\!\! t_P = \sqrt{\hbar G/c^5}
 \approx 5.39120\times 10^{-44} \:\mathrm{s},
 \nonumber \\
1 \!\!&=&\!\! m_P = \sqrt{\hbar c/G} \approx 2.17645\times 10^{-5} \:\mathrm{g},
 \nonumber \\
1 \!\!&=&\!\! \mathcal{E}_P = \sqrt{\hbar c^5/G}
  \approx 1.22090\times 10^{28} \:\mathrm{eV}
  \approx 1.96561\times 10^{16} \:\mathrm{ergs} \approx 543 \:\mathrm{kwh},
 \nonumber \\
1 \!\!&=&\!\! T_P = \sqrt{\hbar c^5/(G k^2_\mathrm{Boltzmann})}
  \approx 1.41679\times 10^{32} \:\mathrm{K},
 \nonumber \\
1 \!\!&=&\!\! F_P = c^4/G \approx 1.2103\times 10^{49} \:\mathrm{dynes},
 \nonumber \\
1 \!\!&=&\!\! N_P = l_P^{-3} t_P^{-1} = c^7/(\hbar^2 G^2)
 \approx 4.39333 \times 10^{141} \:\mathrm{cm}^{-3}\:\mathrm{s}^{-1},
 \nonumber \\
1 \!\!&=&\!\! \rho_P = E_P/l_P^3 = c^7/(\hbar G^2)
 \approx 4.6331\times 10^{114} \:\mathrm{ergs/cm}^3
 \approx 0.84960\times 10^{200} \:\mathrm{eV/Mpc}^3,
 \nonumber \\
1 \!\!&=&\!\! Q_P = \sqrt{4\pi\epsilon_0\hbar c}
 \approx 1.87554592\times 10^{-18}
 \:\mathrm{Coulomb} = 5.62274520\times 10^{-9} \:\mathrm{esu},
 \nonumber \\
1 \!\!&=&\!\! R_P = 1/(4\pi\epsilon_0 c) = 29.9792458 \:\mathrm{ohms},
 \nonumber \\
1 \!\!&=&\!\! E_P = \sqrt{c^7/(4\pi\epsilon_0\hbar G^2)}
 \approx 6.4529\times 10^{59} \:\mathrm{V/cm}.
 \label{eq:2}
 \end{eqnarray}
The numbers above were calculated from values given in the Particle Physics
Booklet \cite{PDG}.  When the $\approx$ sign is used, usually the last two
digits of the corresponding number are uncertain, but when the $=$ sign is
used, the result is exact (by definition), such as the Planck resistance $R_P$
in ohms, which is $10^{-9}$ times the speed of light in cgs units, from the
definition of the permeability of free space in mks units as $\mu_0 =
4\pi\times 10^{-7}$ newtons per square ampere.

The results of our calculations depend mainly on just the following three
physical quantities:  the magnitude of the charge of the electron, positron,
and proton, which we shall here denote by $q$ (instead of the usual $e$, in
order to avoid confusing it with the base of the natural logarithms); the mass
of the electron or positron, $m$; and the mass of the sun, $M_\odot$.  For
comparison we also give the mass of the proton, $m_p$, and the mass of the
neutron, $m_n$.  In Planck units, these various quantities are
\begin{eqnarray}
q \!&=&\! \sqrt{\alpha} \ \approx\ 0.08542454312,
 \nonumber \\
m \!&\approx&\! 4.18543 \times 10^{-23},
 \nonumber \\
M_\odot \!&\approx&\! 9.13617 \times 10^{37},
 \nonumber \\
m_p \!&\approx&\! 7.68509 \times 10^{-20},
 \nonumber \\
m_n \!&\approx&\! 7.69568 \times 10^{-20}.
 \label{eq:3}
 \end{eqnarray}
 
We shall find that the pair production process, for a maximally charged core
collapsing to form a solar mass black hole, depends mainly upon the two
large numbers I shall call $A$ and $B$ (and/or on a third rather large number
$C$ that depends logarithmically upon $A$ and $B$ and is independent of the
solar mass $M_\odot$):
\begin{eqnarray}
A \!&\equiv&\! {q^2 m^2 M_\odot^2 \over \pi}
  \equiv {\alpha\over\pi}\left({GM_\odot m\over \hbar c}\right)^2
  \approx 3.39643251\times 10^{28},
 \nonumber \\
B \!&\equiv&\! {q \over 4\pi m^2 M_\odot}
  \equiv {1\over 4\pi}\left({\hbar c \over GM_\odot m}\right)
    \sqrt{q^2\over 4\pi\epsilon_0 G m^2}
  \approx 42475,
 \nonumber \\
C \!&\equiv&\! \ln{(AB^2)} \equiv \ln{\left({q^4\over 16\pi^3 m^2}\right)}
  \equiv \ln{\left[{\alpha\over 16\pi^3}
    \left({q^2\over 4\pi\epsilon_0 G m^2}\right)\right]}
  \approx 87.00843.
 \label{eq:4}
 \end{eqnarray}

$A$ is thus $\pi$ times the fine structure constant $\alpha \equiv
q^2/(4\pi\epsilon_0\hbar c) \approx 0.007297352568 \approx 1/137.03599911$
times the square of the ratio of the Schwarzschild radius of a solar mass black
hole, $2M_\odot = 2GM_\odot/c^2 \approx 295\,325.008$ cm, to the Compton
wavelength of an electron, $\lambda = 2\pi/m = 2\pi\hbar/(mc) \approx
2.426310238\times 10^{-10}$ cm.  As we shall see, the largeness of $A$ is the
key to how much weaker than dyadosphere values are the maximum electric field
and the maximum pair production rate that can develop from a collapsing charged
core.  (The experimental determination of $A$ does not depend upon the
relatively large uncertainty in Newton's constant $G$ in cgs units, because it
is $GM_\odot$ rather than $M_\odot$ that is determined directly by Solar System
measurements, so the value of $A$ is known much more precisely than quantities
like $B$ that do involve the experimental determination of $G$.)

If we use the well-known fact that in Planck units, stellar masses (particularly
maximum white dwarf and neutron star masses) are of the order of the inverse
square of the proton mass,
 \begin{equation}
 \tilde{M} \equiv m_p^{-2} \approx 1.69318\times 10^{38}
  \approx 1.85327 M_\odot
 \label{eq:c}
 \end{equation}
(which coincidentally is within half of one percent of the largest prime found
without computers,
$2^{127}-1=170\,141\,183\,460\,469\,231\,731\,687\,303\,715\,884\,105\,727$,
which might be useful mnemonically to some readers), then one may give crude
approximations for $A$ and for $B$ that just depend on the elementary charge
$q$ and on the masses $m$ and $m_p$ of the electron and proton respectively:
\begin{eqnarray}
\tilde{A} \!&\equiv&\! {q^2 m^2 \tilde{M} \over \pi}
  \equiv {\alpha\over\pi}\left({m\over m_p}\right)^2
                    \left({\hbar c\over G m_p^2}\right)
  \equiv {\alpha\over\pi}\left({m\over m_p}\right)^2 \tilde{M}
  \approx 1.16654\times 10^{29},
 \nonumber \\
\tilde{B} \!&\equiv&\! {q \over 4\pi m^2 \tilde{M}}
  \equiv {\sqrt{\alpha}\over 4\pi} \left({m_p\over m}\right)^2
  \approx 22919.
 \label{eq:4b}
 \end{eqnarray}
Thus $\tilde{A}$, and hence $A$, is very roughly a solar mass in Planck units,
reduced by one power of the fine structure constant $\alpha$ and by two powers
of the ratio $m/m_p$ of the electron mass to the proton mass.  Similarly,
$\tilde{B}$, and hence roughly $B$, is mainly the square of the ratio $m_p/m$
of the proton mass to the electron mass, but reduced by the square root of the
fine structure constant and by the ratio of the area of a unit square to the
area of a unit sphere.

As one may derive from the definition below of a dyadosphere, an extreme
charged Reissner-Nordstrom black hole ($Q=M$) has a dyadosphere if its mass is
less than the critical mass
 \begin{equation}
 M_c \equiv q/m^2 = 4\pi B M_\odot \approx 5.33751\times 10^5 M_\odot,
 \label{eq:5}
 \end{equation}
so $B$ is $1/4\pi$ times the ratio of the critical mass $M_c$ to the solar mass
$M_\odot$.  Later we shall see that it is impossible to form a black hole with
$Q/M \stackrel{>}{\sim} 0.99$ if $M < BCM_\odot \approx 3\,695\,650 M_\odot
\sim 3.7\times 10^6 M_\odot$.

The combination $C$ of the two independent large numbers $A$ and $B$ depends
only on the properties of the electron and is the logarithm of the product of
the reciprocal of $16\pi^3$, the fine structure constant $\alpha \equiv q^2
\equiv q^2/(4\pi\epsilon_0\hbar c)$, and the square of the magnitude $q/m
\equiv (q/m)/\sqrt{4\pi\epsilon_0 G}$ of the charge-to-mass ratio of the
electron.  $C$ is large because the square of the magnitude of the
charge-to-mass ratio of the electron, the ratio of the electrostatic repulsion
to the gravitational attraction between two electrons, is very large:  $(q/m)^2
\approx 4.1657\times 10^{42}$.

\subsection{Pair production rate and definition of a dyadosphere}

The Schwinger pair production \cite{Sch} gives a rate $\mathcal{N}$ of
electron-positron pairs per four-volume (per three-volume and per time) in a
uniform electromagnetic field that is \cite{Nik}
 \begin{equation}
 \mathcal{N} = {(qE)(qB)\over(2\pi)^2}\coth{\left({\pi B\over E}\right)}
                 \exp{\left(-{\pi m^2\over qE}\right)},
 \label{eq:6}
 \end{equation}
where $E$ and $B$ are the electric and magnetic field magnitudes in a frame in
which the electric and magnetic fields are parallel (which is a frame in which
the electric field has the minimum value).

Note that the exact one-loop pair production rate given above is not quite the
same as twice the imaginary part of the one-loop effective action per
four-volume, which is given by an infinite series \cite{Sch,Nik}, of which
the pair production rate is only the first term.  However, for weak fields the
first term dominates, so in such cases it is a good approximation to say that
the pair production rate is twice the imaginary part of the one-loop effective
action per four-volume.

Here we shall be mainly interested in the case in which there is no magnetic
field, in which case the pair production rate may be obtained from the formula
above by taking the limit that $B/E$ goes to zero:
 \begin{equation}
 \mathcal{N} = {q^2 E^2\over 4\pi^3}\exp{\left(-{\pi m^2\over qE}\right)}
             \equiv {m^4\over 4\pi} {e^{-w}\over w^2},
 \label{eq:7}
 \end{equation}
where
 \begin{equation}
 w \equiv {\pi m^2\over qE} \equiv {\pi E_c \over E},
 \label{eq:8}
 \end{equation}
if one defines the critical electric field strength $E_c$ to be
 \begin{equation}
 E_c \equiv {m^2 \over q} \equiv {m^2 c^3 \over \hbar q}
  \approx 2.05068\times 10^{-42}
   \approx 1.32328539 \times 10^{16} \:\mathrm{V/cm}.
 \label{eq:9}
 \end{equation}

Ruffini \cite{Ruf98,PRX,RSWX,RX,Ruf} has defined a {\it dyadosphere} to be a
region of spacetime in which the electric field $E$ is greater than the
critical electric field $E_c$.  Therefore, it is a region in which $w < w_c =
\pi$, and a region in which the electron-positron pair production rate is
 \begin{eqnarray}
 \mathcal{N} > \mathcal{N}_c \!&=&\! {m^4\over 4\pi} {e^{-\pi}\over \pi^2}
  \approx 0.0003484287930 \: m^4
   \approx 1.06924\times 10^{-93} \\ \nonumber
    &\approx& 4.69752273\times 10^{48}\:\mathrm{cm}^{-3}\:\mathrm{s}^{-1}
     \approx 7.449533923\times 10^{58} M_\odot^{-4}.
 \label{eq:10}
 \end{eqnarray}

The last number means that in a cube of edge length $M_\odot \approx
1.47662504$ kilometers, half the Schwarzschild radius of the sun, and in a time
$M_\odot \approx 4.92549095$ microseconds, which would form a four-volume of
$M_\odot^4 \approx 1.58584308 \times 10^{10} \:\mathrm{cm}^3 \:\mathrm{s}$ that
might be regarded as a minimum for typical macroscopic astrophysical
four-volumes (about the same four-volume as a liter multiplied by half a year),
more than $7\times 10^{58}$ pairs would be produced in a dyadosphere (under the
highly hypothetical situation in which the dyadosphere could be maintained over
this spacetime region, and also ignoring the effects of the pairs produced, as
well as spacetime curvature).

To put it another way, in a volume contained within a sphere in flat space of
the Schwarzschild radius of the sun, $16\pi M_\odot^3/3 \approx 53.9460048$
cubic kilometers, the dyadosphere pair production rate would be more than
$2.5\times 10^{65}$ pairs produced per second.  For the positrons produced to
have a total charge equal to a solar mass (in Planck units, as always when the
units are not otherwise specified), one would need $M_\odot/q \approx
1.06950\times 10^{39}$ positrons produced, and the time it would take for this
would be less than $10^{-26}$ seconds (again making the hypothetical assumption
that the dyadosphere could be maintained over this volume for this length of
time).

This shows that the minimum dyadosphere pair production rate, though only $4\pi
e^{-\pi} \approx 0.5430421124$ in units of the electron Compton wavelength
\newline $2\pi/m \approx 2.42631024 \times 10^{-10}$ cm and the corresponding
time of $8.093299792 \times 10^{-21}$ seconds, is utterly enormous at
astrophysical scales.  This strongly suggests that dyadospheres will never form
over macroscopic astrophysical scales.  Indeed, the calculations below confirm
this and show that under extremely conservative assumptions, the pair
production rate from astrophysical gravitational collapse is always less than
$10^{-26}$ of that of a dyadosphere.  (Under the plausible but not rigorous
arguments of the Introduction that the charge is ejected when the electrostatic
repulsion exceeds the gravitational attraction, the upper limit in a region of
the order of the size of a black hole would be trillions of orders of magnitude
weaker than a dyadosphere, a factor of more than $10^{10^{13}}$, and hence
completely negligible.)

\subsection{Crude upper limit of pair production rate}

One can make the following crude estimate of the upper limit of the pair
production rate from any macroscopic astrophysical process that has length
scale $L$ and time scale $T$.

Suppose the electric field is
 \begin{equation}
 E \equiv {\pi E_c\over w} \equiv {\pi m^2\over q w} = {\tilde{Q}\over L^2},
 \label{eq:d}
 \end{equation}
thus determining a characteristic effective charge $\tilde{Q}$ for the field
producing the pairs at the rate per 4-volume given by Eq. (\ref{eq:7}),
 \begin{equation}
 \mathcal{N} = {\pi^2 e^{\pi-w} \over w^2} \mathcal{N}_c
 \approx 1.07286 \times 10^{57} \, {e^{-w} \over w^2}
 \:\mathrm{cm}^{-3}\:\mathrm{s}^{-1}.
 \label{eq:e}
 \end{equation}
This gives
 \begin{equation}
 w + 2\ln{w} = \ln{\left(\pi^2 e^\pi {\mathcal{N}_c\over\mathcal{N}}\right)} > 1
 \label{eq:f}
 \end{equation}
for $\mathcal{N} < \pi^2 e^\pi \mathcal{N}_c = m^4/(4\pi)$, which will be
assumed for now and confirmed later by Eq. (\ref{eq:i}) below for $T
\,\stackrel{>}{\sim}\, L \,\stackrel{>}{\sim}\, 4.3\times 10^{-8}
\mathrm{cm}$.  Thus Eq. (\ref{eq:d}) gives $\tilde{Q} < \pi m^2 L^2/q$.

For the timescale of the electric field to be at least $T$, the charge
$\tilde{Q}'$ produced in this same time, which tends to discharge the electric
field, cannot greatly exceed $\tilde{Q}$.  This then gives
 \begin{equation}
 \tilde{Q}' \sim q \mathcal{N} L^3 T
  = {\mathcal{N}\over\mathcal{N}_c}{q m^4 e^{-\pi}\over 4\pi^3} L^3 T
  \,\stackrel{<}{\sim}\, \tilde{Q} < {\pi m^2 L^2 \over q},
 \label{eq:g}
 \end{equation}
or
 \begin{equation}
 {\mathcal{N}\over\mathcal{N}_c} \,\stackrel{<}{\sim}\,
 {4\pi^4 e^\pi\over q^2 m^2 L T}
 \approx 8.45\times 10^{-26} \, {M_\odot^2\over L T},
 \label{eq:h}
 \end{equation}
which is enormously smaller than unity for macroscopic astrophysical length and
time scales $L$ and $T$.

This strongly argues that no matter what the situation, if the time and length
scales obey

 \begin{equation}
 T \,\stackrel{>}{\sim}\, L
  \gg {2\pi^2 e^{\pi/2}\over qm} \approx 2.66\times 10^{25}
  \approx 4.3\times 10^{-8} \mathrm{cm},
 \label{eq:i}
 \end{equation}
then dyadospheres of such a macroscopic size will not form.

In the example below, we shall find a more precise upper bound on the pair
production rate in the case of spherical symmetry that is slightly more than
one order of magnitude stronger.  This example confirms the limit above and
strengthens the evidence that pair production rates in macroscopic processes
must be many, many orders of magnitude below dyadosphere rates.

\subsection{Approximate estimate of the discharge rate}

We consider a positively charged spherical core of mass $M$ freely collapsing
into a black hole, with the surface at a radius $R(t)$, and with vacuum outside
(except for the electromagnetic field and pairs produced by it, which we assume
will not significantly modify the Schwarzschild geometry; including such
modifications would reduce the electric field even further than the
conservative upper limits we shall find below).  Outside the surface of the
core, pairs will be produced, with the positrons moving outward and the
electrons moving inward.  (Since the core is expected to be highly conducting,
virtually all of its excess charge would be at the surface, so inside the core
there would be a negligible macroscopic field and hence negligible pair
production.)  We shall confirm in Section 4 that the interactions between
individual positrons and electrons (e.g., the annihilation probability for each
particle) is utterly negligible.

As the electrons pass inward through the outer boundary of the core over time,
they reduce the value of the charge of the core, $Q(t)$, limiting the value of
the electric field outside, $E(t,r) \approx Q(t)/r^2$, under the assumption
(which can be verified from the results) that at any one time the charge
contributed by the pairs outside the core radius $R(t)$ is a small fraction of
$Q(t)$, so long as one does not go to such huge radii that it includes the
outgoing positrons emitted over a large fraction of the previous infall of the
core.  (At such huge radii where the charge may differ significantly from that
of the core at the same time, because of the outgoing positrons in between, the
inverse-square falloff of the electric field would make it so weak that pair
production would be negligible, so we need not consider such regions.)

We also make the assumptions, which will be verified later (at least for $M \ll
10^6 M_\odot$) that $Q(t) \ll M$ once the core gets within a few orders of
magnitude of the Schwarzschild radius $2M$, that the total energy that goes
into the pairs is also much smaller than $M$ (so that the core mass $M$ stays
very nearly constant), and that (with no restriction on $M$) the probability
for any one of the particles produced to annihilate with an antiparticle is
negligible.

Because the pair production rate per four-volume, $\mathcal{N} =
(m^4/4\pi)w^{-2}e^{-w}$, decreases exponentially rapidly with $w(t,r) = \pi
m^2/qE \approx \pi m^2 r^2/qQ(t)$, and because the pairs produced decrease
$Q(t)$ and hence increase $w(t,r)$ at fixed $r$, there will be a
self-regulation of $w(t,r)$ (described more precisely by the differential
equations of the following subsections, though we do not need this for the
approximate estimate of this subsection).  In particular, if we define
 \begin{equation}
 z(t) \equiv w(t,R(t)) = {\pi E_c\over E(R(t))} = {\pi m^2 R(t)^2 \over q Q(t)},
 \label{eq:11}
 \end{equation}
which gives the pair production rate $\mathcal{N}(t) = (m^4/4\pi)z^{-2}e^{-z}$
at the surface of the core (where the rate is maximal at that time, since the
electric field has an inverse-$r^2$ falloff at greater radii $r>R$), the
self-regulation will keep $z(t)$ changing only slowly with $t$ and $R(t)$. 
Hence $Q(t)$ will vary roughly in proportion to $R(t)^2$.

This means that the logarithmic rate of change of $Q(t)$ will be roughly twice
the logarithmic rate of change of $R(t)$.  If we do a Newtonian analysis of
free fall from rest at infinity, we find that $dR/dt = -\sqrt{2M/R}$, which
makes the logarithmic rate of change of $R(t)$ equal to $-1/(2M)$ times
$(2M/R)^{3/2}$.  As the core approaches the horizon at $R=2M$, this Newtonian
estimate of the logarithmic rate will approach $-1/(2M)$, so the logarithmic
rate of change of $Q(t)$ will approximately approach $-1/M$.  (Here we are
ignoring relativistic corrections, but it turns out that they make only a
relatively small difference in the maximum electric field.)

Now we can calculate $dQ/dt$ from the pair production rate (as a function of
$Q$) and set it equal to $-Q/M$ when $R \sim 2M$ to solve for $Q$ and hence the
pair production rate.  Because the logarithmic rate of change of $Q$ is
enormously less than that given above for a dyadosphere, the pair production
rate must be suppressed by rather large values of $w$ outside the core.  Since
$w \approx \pi m^2 r^2/qQ$ increases proportional to the square of the radius,
the suppression will rather rapidly increase with the radius outside the core
(which we are now taking to be at $R \sim 2M$).  This means that the pair
production rate will decrease roughly exponentially with the radius and will
reach a value a factor of $1/e$ smaller than at the core surface itself (where
$w=z = \pi m^2 R^2/qQ \sim 4\pi m^2 M^2/qQ$) roughly when $w$ increases by
unity.  Since the logarithmic rate of increase of $w$ with $r$ at $r=R$ is
$2/R$, the point at which the pair production rate will have dropped by the
factor of $1/e$ will be at $r-R \approx R/(2z) \ll R$, the last inequality
coming because $z \gg 1$ in order that the pair production rate be much less
than dyadosphere rates.

With a roughly exponential decrease in the pair production rate with radius, at
a logarithmic rate greater by a factor of $z/2 \gg 1$ than the logarithmic rate
at which the radius grows, the total pair production rate per time is roughly
the pair production rate per four-volume at the surface of the core
($\mathcal{N}(R) = (m^4/4\pi) z^{-2} e^{-z}$), multiplied by the effective
volume where most of the pair production is occurring, which in this case is
roughly the area $4\pi R^2$ of the core surface (or of the black hole formed by
the collapsing core) multiplied by the radial distance $r-R \approx R/(2z)$ out
to where the pair production rate per four-volume has decreased by a factor of
$1/e$.  (For the pair production rate per coordinate time $t$, the relativistic
correction that makes the proper radial distance greater than $r-R$ is
compensated in the approximately Schwarzschild metric outside the core by the
relativistic correction that makes the proper time smaller than $\Delta t$ by
what is precisely the same factor in the Schwarzschild metric.)

When the number rate per time is multiplied by the charge $-q$ of each ingoing
electron, one gets at $R \sim 2M$ that
 \begin{equation}
 {dQ\over dt} \approx -q \mathcal{N}(R) 4\pi R^2 {R\over 2z}
  \sim {-4 q m^4 M^3 \over z^3 e^z}.
 \label{eq:12}
 \end{equation}
For the logarithmic rate of change of $Q$ to be roughly $-1/M$, we set $dQ/dt$
equal to $-Q/M = -\pi m^2 R^2/(qMz) \sim - 4\pi m^2 M/(qz)$.  This leads to the
equation for $z$,
 \begin{equation}
 z^2 e^z \sim Z \equiv {q^2 m^2 M^2 \over \pi}
   \equiv A \left({M\over M_\odot}\right)^2
   \equiv A \mu^2 \approx 3.39643251\times 10^{28} \mu^2,
 \label{eq:13}
 \end{equation}
where
 \begin{equation}
 \mu \equiv {M\over M_\odot}
 \label{eq:14}
 \end{equation}
is the ratio of the mass $M$ of the freely collapsing core to the mass $M_\odot$
of the sun.

Since $A$ is so large, for $|2\ln{\mu}| \ll \ln{A}$ one gets the crude explicit
solution
 \begin{equation}
 z \sim \ln{A} -2\ln{\ln{A}} +2\ln{\mu} \approx 57.325 +2\ln{\mu}.
 \label{eq:15}
 \end{equation}
One can also solve Eq. (\ref{eq:13}) numerically for $z$ when $\mu=1$
($M=M_\odot$) to get what I shall call $z_*$:
 \begin{equation}
 z_* \approx 57.588464.
 \label{eq:16}
 \end{equation}
Then for any $|2\ln{\mu}| \ll \ln{A}$, one gets the approximate solution
 \begin{equation}
 z \approx z_* + {2 z_* \over z_* + 2} \ln{\mu}
  \approx 57.588464 + 1.932873 \ln{\mu}.
 \label{eq:17}
 \end{equation}
 
This means that the ratio of the electric field value $E(2M)$ at the surface of
the collapsing core (when it enters the black hole) to the critical electric
field value $E_c$ for the definition of a dyadosphere is
 \begin{equation}
  {E(2M)\over E_c} = {\pi\over z} \approx {\pi\over z_*} - {2\pi \ln{\mu}
   \over z_*(z_* + 2)} \approx 0.054552465 - 0.001830974 \ln{\mu}.
 \label{eq:18}
\end{equation} 

Thus this approximate estimate would indicate that the electric field outside a
charged collapsing core is always less than about 5.5\% of that of a
dyadosphere, differing from that of a dyadosphere by a factor of at least 18. 
(We shall see later that the numerical solution of the ordinary differential
equation governing the charge during the collapse, including relativistic
effects, agrees with the result above to about three decimal places, so the
result above is quite accurate.)

From this solution, one can see that the pair production rate at the surface of
the core when it crosses the horizon at $R=2M$ is
 \begin{equation}
 \mathcal{N}(R) \sim {m^4 \over 4\pi} z_*^{-2} e^{-z_*}
  \sim {m^4 \over 4\pi A\mu^2} = {\pi^2 e^{\pi}\over A\mu^2} \mathcal{N}_c
  = {m^2\over 4 q^2 M^2}
   \sim 7 \times 10^{-27} \mu^{-2} \mathcal{N}_c,
 \label{eq:19}
 \end{equation}
which for $M > M_\odot$ or $\mu \equiv M/M_\odot > 1$ is more than 26 orders of
magnitude smaller than the minimum pair production rate $\mathcal{N}_c$ of a
dyadosphere.

Therefore, the approximate algebraic estimate is that the pair production rate
is never more than about $10^{-26}$ times the minimum amount for a
dyadosphere.  In the next subsections we shall indeed confirm from the
solutions of differential equations that this algebraic estimate is indeed a
good approximation for the maximum electric field.  This result implies that it
is very unlikely that dyadospheres can form from the collapse of charged cores,
even if somehow all discharge mechanisms are eliminated other than the pair
production itself.

\subsection{Differential equations for the discharge}

In this subsection we shall derive and in the next two subsections solve
(approximately) the differential equations for the discharge of the collapsing
core.  The differential equations will be derived under the assumption that the
tunneling distance for a pair to come into real existence,
$L_\mathrm{tunnel}\sim m/(qE) = z/(\pi m) \sim 7\times 10^{-10}$ cm, is much
less than the astrophysical length scales for the collapsing core, which is a
very good approximation.  The differential equations will be solved under the
approximation that this tunneling length is also significantly greater than the
Compton wavelength of an electron, which implies that $w \equiv \pi m^2/(qE)
\gg 1$ and hence that the pair production is mostly confined to a radial region
$r-R \sim R/w \ll R$ that is much smaller than the radius $R$ of the surface of
the charged collapsing core.  This latter approximation is less good but still
leads to errors of only a few percent.

After the charged particles are produced in pairs, they will be accelerated by
the radial electric field (electrons inward and positrons outward, under the
assumption here that the core is positively charged).  Each time a charged
particle travels a distance $m/(qE)$ parallel to the electric field, it will
gain an additional kinetic energy equal to its rest mass.  Thus it will very
quickly accelerate to a huge gamma factor and move very nearly at the speed of
light.  As a result, one will get effectively a null 4-vector of positive
current density (highly relativistic positrons) moving radially outward and a
null 4-vector of negative current density (highly relativistic electrons)
moving radially inward.

It is most convenient to describe this current in terms of radial null
coordinates, say $U$ and $V$, so that the approximately Schwarzschild metric
outside the collapsing core may be written as
 \begin{equation}
 ds^2 = - e^{2\sigma} dU dV + r^2(U,V)(d\theta^2 + \sin^2{\theta} d\phi^2).
 \label{eq:20}
 \end{equation}
For example, if one defines the radial variables $u$ and $v$ (not to be confused
with the null coordinates $U$ and $V$ above) and $r_*$ by
 \begin{equation}
 u \equiv v^2 \equiv {2M\over r},
 \label{eq:21}
 \end{equation}
 \begin{equation}
 r_* \equiv \int {dr\over 1-u}
  = 2M \left[ {1\over u} + \ln{\left({1\over u}-1\right)} \right],
 \label{eq:22}
 \end{equation}
then a common choice of null coordinates for the exterior region of the
Schwarzschild metric is
 \begin{eqnarray}
 \tilde{U} \!&=&\! t - r_* \nonumber \\
   \!&=&\! t - 2M \left[ {1\over u} + \ln{\left({1\over u}-1\right)} \right],
   \nonumber \\
 \tilde{V} \!&=&\! t + r_* \nonumber \\
   \!&=&\! t + 2M \left[ {1\over u} + \ln{\left({1\over u}-1\right)} \right].
 \label{eq:23}
 \end{eqnarray}
With this choice of null coordinates, one gets $e^{2\tilde{\sigma}} = 1-u =
1-v^2$.  Other choices of radial null coordinates would be any smooth monotonic
function of $\tilde{U}$ as the retarded radial null coordinate, and any smooth
monotonic function of $\tilde{V}$ as the advanced radial null coordinate. 
Below we shall give another choice of null coordinates $(U,V)$ that is useful
for getting an approximation to the discharge rate.

Now we can write the nearly-null outward number flux 4-vector of positrons as
$\mathbf{n_+} = n_+^V \partial_V$ and the nearly-null inward number flux
4-vector of electrons as $\mathbf{n_-} = n_-^U \partial_U$.  Since each
positron has charge $q$ and each electron has charge $-q$, the total current
density 4-vector is
 \begin{equation}
 \mathbf{j} = q \mathbf{n_+} - q \mathbf{n_-}
  = q n_+^V \partial_V - q n_-^U \partial_U.
 \label{eq:24}
 \end{equation}

The radial electric field of magnitude $E=Q/r^2$ has the electromagnetic field
tensor
 \begin{equation}
 \mathbf{F} = -{1\over 2}E e^{2\sigma} dU \wedge dV
  = -{Q\over 2 r^2} e^{2\sigma} dU \wedge dV,
 \label{eq:25}
 \end{equation}
where $Q=Q(U,V)$ is the charge inside the sphere labeled by $(U,V)$ (and which
is a function only of these two null coordinates, because of the assumed
spherical symmetry).

Then from Maxwell's equations (essentially just Gauss's law here), we may
deduce that the null components of the current density vector are
 \begin{eqnarray}
 j^V \!&=&\! q n_+^V = {-2e^{-2\sigma} Q_{,U} \over 4\pi r^2}
     = {+Q^{,V} \over 4\pi r^2},
   \nonumber \\
 j^U \!&=&\! -q n_-^U = {+2e^{-2\sigma} Q_{,V} \over 4\pi r^2}
     = {-Q^{,U} \over 4\pi r^2}.
 \label{eq:26}
 \end{eqnarray}
 
Although of course the current density 4-vector field is conserved, the number
flux 4-vectors $\mathbf{n_+}$ and $\mathbf{n_-}$ of the positrons and electrons
are not.  Their 4-divergences are each equal to the pair production rate
$\mathcal{N}$ when we can neglect annihilations, as we can here with the
density of pairs being sufficiently small.  When these 4-divergences are
written in terms of the charge $Q(U,V)$, one gets the following partial
differential equation for the pair production and discharge process:
 \begin{equation}
 Q_{,UV} = -2\pi q r^2 e^{2\sigma} \mathcal{N}
  =-{q^3 Q^2 e^{2\sigma}\over 2\pi^2 r^2}
      \exp{\left(-{\pi m^2 r^2\over qQ}\right)}.
 \label{eq:27}
 \end{equation}

In covariant notation, with $^2\Box Q = -4e^{-2\sigma}Q_{,UV}$ being the
covariant Laplacian in the 2-dimensional metric $^2ds^2=-e^{2\sigma}dU dV$ and
with $\Box$ being the covariant Laplacian in the full 4-dimensional metric, the
differential equations for the pair production and discharge may be written as
 \begin{eqnarray}
 8\pi q r^2 \mathcal{N}
  \!&=&\! {2 q^3 Q^2\over \pi^2 r^2}\exp{\left(-{\pi m^2 r^2\over qQ}\right)}
  \nonumber \\
  \!&=&\! ^2\Box Q
  = \Box Q - {2\over r}\nabla r \cdot \nabla Q
  \nonumber \\
  \!&=&\! r\Box\left({Q\over r}\right) - rQ\Box\left({1\over r}\right)
  = r\Box\left(rE\right) - r^3 E\Box\left({1\over r}\right).
 \label{eq:28}
 \end{eqnarray}
or, explicitly in terms of the electric field $E = Q/r^2$, as
 \begin{equation}
 8\pi q r^2 \mathcal{N}
  = {2q^3 r^2 E^2\over \pi^2} \exp{\left(-{\pi m^2\over qE}\right)}
  =\  ^2\!\Box (r^2E)
  = r\Box\left(rE\right) - r^3 E\Box\left({1\over r}\right).
 \label{eq:29}
 \end{equation}

Since for the sake of argument in this section we assume that the positively
charged particles at the surface of the collapsing core do not escape to the
outside, and since there is no electric field inside to produce particles
there, the boundary condition at the surface of the collapsing core is that
there is no outward flux of positrons there, so that $Q_{,U} = 0$ at the core
surface.  The boundary condition at infinite radii is that we assume that there
are no incoming electrons there, but only outgoing positrons pair-produced by
the electric field at finite radius, so at radial infinity, $Q_{,V} = 0$.  The
boundary condition in the infinite past is that we assume that $Q$ is as large
as it can be and still have the core collapse gravitationally, so there we set
$Q = M$ but then make the idealized assumption of ignoring the effect of the
electrostatic repulsion on the collapse of the core.

For an analysis of the solution of the partial differential equation
(\ref{eq:27}) or (\ref{eq:28}) with the boundary conditions given above and
with the surface of the collapsing core freely falling inward from infinity in
an assumed external Schwarzschild geometry (ignoring the gravitational effect
of the electric field, which will be negligible once the core at large radius
discharges enough to get its charge $Q$ to become much less than its mass $M$),
it is useful to shift from the commonly used null coordinates
$(\tilde{U},\tilde{V})$, given by Eq. (\ref{eq:23}) above, to new null
coordinates $(U,V)$ which on the core surface are both set equal to the infall
velocity of the core surface (as seen in an orthonormal frame carried by a
static observer), $v \equiv \sqrt{2M/r}$, which runs from 0, at the idealized
beginning of the core surface collapse at radial infinity, to 1 when the core
surface crosses the event horizon at the Schwarzschild radius $r=2M$ and enters
the black hole.  (The variable $v$, and hence $U$ and $V$ at the core surface,
would increase to infinity at the black hole singularity at $r=0$, but since
nothing inside the black hole can be visible to the outside, we shall ignore
what happens inside and just consider the values of the electric field outside,
where $v<1$.)

One can readily calculate that the freely collapsing core surface, moving along
radial geodesics in the Schwarzschild metric with unit conserved
energy-at-infinity per rest mass, has Schwarzschild coordinates
 \begin{eqnarray}
 t &=& 2M[-{2\over 3}v^{-3} - 2v^{-1} - \ln{(1-v)} + \ln(1+v)],
 \nonumber \\
 r &=& {2M\over v^2},
 \label{eq:30}
 \end{eqnarray}
where we have chosen the origin of the $t$ coordinate to avoid the explicit
appearance of a constant of integration.
The tortoise radial coordinate $r_*$ and the proper time $\tau$ along the
worldlines of the core surface have the form
 \begin{eqnarray}
 r_* &=& 2M[v^{-2} - 2\ln{v} + \ln{(1-v)} + \ln(1+v)],
 \nonumber \\
 \tau &=& -{4M\over 3v^3},
 \label{eq:31}
 \end{eqnarray}
where we have chosen the origin of the proper time to be the proper time at
which the core surface hits the singularity at $r=0$ inside the black hole
(so that $\tau$ is then negative for all the physical times before then).

Therefore, at the core surface at each value of $v$, Eq. (\ref{eq:23}) gives
the values of the null coordinates above as
 \begin{eqnarray}
 \tilde{U} \!&=&\! 4M[- {1\over 3}v^{-3} - {1\over 2}v^{-2} - v^{-1}
                      + \ln{v} - \ln{(1-v)}],
   \nonumber \\
 \tilde{V} \!&=&\! 4M[- {1\over 3}v^{-3} + {1\over 2}v^{-2} - v^{-1}
                      - \ln{v} + \ln{(1+v)}].
 \label{eq:32}
 \end{eqnarray}

Now we want new null coordinates $U=U(\tilde{U})$ and $V=V(\tilde{V})$
both to equal $v$ along the collapsing core surface, so the inverse relations
must be
 \begin{eqnarray}
 \tilde{U}(U) \!&=&\! 4M[- {1\over 3}U^{-3} - {1\over 2}U^{-2} - U^{-1}
                      + \ln{U} - \ln{(1-U)}],
   \nonumber \\
 \tilde{V}(V) \!&=&\! 4M[- {1\over 3}V^{-3} + {1\over 2}V^{-2} - V^{-1}
                      - \ln{V} + \ln{(1+V)}].
 \label{eq:33}
 \end{eqnarray}
The differential forms of these relations are
 \begin{eqnarray}
 d\tilde{U} \!&=&\! {4M dU \over U^4(1-U)},
   \nonumber \\
 d\tilde{V} \!&=&\! {4M dV \over V^4(1+V)}.
 \label{eq:34}
 \end{eqnarray}
Unfortunately, it seems unlikely that the direct relations $U=U(\tilde{U})$ and
$V=V(\tilde{V})$ can be written explicitly in closed form in terms of
elementary functions.

In terms of these null coordinates $(U,V)$, the Schwarzschild metric
Eq. (\ref{eq:20}) becomes
 \begin{equation}
 ds^2 = - {16 M^2 (1-v^2) dU dV \over U^4(1-U) V^4(1+V)}
 + {4 M^2 \over v^4} (d\theta^2 + \sin^2{\theta} d\phi^2),
 \label{eq:35}
 \end{equation}
where $v(U,V)$ is defined implicitly by
 \begin{eqnarray}
 &v&\!\!\!\!\!^{-2} - 2\ln{v} + \ln{(1-v^2)} = {r_*\over 2M} 
 = {\tilde{V} - \tilde{U} \over 4M}
   \nonumber \\
 &=& {1\over 3}U^{-3} - {1\over 3}V^{-3} + {1\over 2}U^{-2}
      +{1\over 2}V^{-2} + U^{-1} - V^{-1}
   \nonumber \\
   &&-\ln{U} - \ln{V} + \ln{(1-U)} + \ln{(1+V)}.
 \label{eq:36}
 \end{eqnarray}

It is convenient to define new time and radial coordinates
 \begin{eqnarray}
 T \!&=&\! {U+V \over 2},
   \nonumber \\
 X \!&=&\! {V-U \over 2}.
 \label{eq:37}
 \end{eqnarray}
Then on the collapsing core surface, $T = v = \sqrt{2M/r} =
[4M/(-3\tau)]^{1/3}$ and $X=0$, with $-\tau$ being the proper time remaining
until the surface reaches the curvature singularity at $r=0$ (where $T$
increases to infinity, though $T$ increases only from 0, at $\tau = -\infty$
where the idealized freely falling core surface has $r=\infty$ and zero
velocity, to 1 before the collapsing core surface enters the black hole horizon
at the Schwarzschild radius $r=2M$ and the inward radial velocity $v$
approaches the speed of light with respect to a sequence of static observers
outside the black hole).  $X$ rises above 0 as one moves radially outward from
the collapsing core surface.

Nearly all of the pair production will occur close to the collapsing core
surface (since the inverse-square falloff of the electric field with $r$ will
make the pair-production rate drop rapidly with $r$), where $X \ll T^2 < T$. 
Thus we can express the metric functions exactly as functions of $T$ but for
$X \ll T^2 < T$ only as an expansion in $X/T^2$ and in $X/T$.
To second order in $X/T^2$ and in $X/T$, Eq. (\ref{eq:36}) leads to
 \begin{equation}
 v \approx T - {X\over T} + {3-5T^2\over 2T^3}X^2
  = {U+V\over 2} - {V-U\over U+V} + {12-5(U+V)^2\over 4(U+V)^3}(V-U)^2.
 \label{eq:38}
 \end{equation}
Inserting this back into the metric Eq. (\ref{eq:35}), and also substituting
$U=T-X$ and $V=T+X$, then gives the Schwarzschild metric near the freely
collapsing core surface, to second order in $X/T$, as
 \begin{eqnarray}
 ds^2 \approx &-& {16 M^2 \over T^8}\left(1+{4X^2\over T^2}\right)(-dT^2+dX^2)
 \nonumber \\
 &+& {4 M^2 \over T^4}\left(1+{4X\over T^2}+{4+10T^2\over T^4}X^2\right)
     (d\theta^2 + \sin^2{\theta} d\phi^2).
 \label{eq:39}
 \end{eqnarray}

For a good estimate of the dominant part of the pair-production rate and of the
maximum value attained by the electric field outside the black hole, it seems
to suffice to use this metric Eq. (\ref{eq:39}) simply to first order in $X$
(which, when small, is proportional to the distance from the collapsing core
surface, though with a $T$-dependent proportionality factor, so that the
physical distance, in a radial spacetime direction orthogonal to the radially
infalling geodesic worldlines along the collapsing core surface, is roughly
$4M/T^4$ times $X$ for sufficiently small $X/T$).  If we return to the null
coordinates $(U,V)$ but keep only the single first-order term in $2X=V-U$,
we get
 \begin{equation}
 ds^2 \approx - {2^{12} M^2\over (U+V)^8} dU dV
 + {2^6 M^2 \over (U+V)^4} \left[ 1 + {8(V-U)\over(U+V)^2} \right]
      (d\theta^2 + \sin^2{\theta} d\phi^2).
 \label{eq:40}
 \end{equation}

\subsection{Approximate solution for the discharge}

In this subsection we shall construct an approximate solution of the partial
differential equation (\ref{eq:27}) for the pair production and discharge
process.

To simplify the constants appearing in the various equations, it is useful for
each core mass $M = M_\odot \mu$ to define the normalized charge
 \begin{equation}
 y \equiv {q^3 Q \over 4 \pi^2 Z} \equiv {qQ\over 4\pi m^2 M^2},
 \label{eq:41}
 \end{equation}
where $q$ and $m$ are the charge and mass of the positron (in Planck units) and
where $Z \equiv q^2 m^2 M^2/\pi \equiv A\mu^2$ was defined in Eq. (\ref{eq:13}),
with $A \equiv q^2 m^2 M_\odot^2/\pi \approx 3.4 \times 10^{28}$ being defined
in Eq. (\ref{eq:4}) and $\mu$ being defined in Eq. (\ref{eq:14}) as the ratio
of the core mass $M$ to the solar mass $M_\odot$.

It is also useful to use the quantity $w \equiv \pi E_c/E \equiv \pi m^2/(qE) =
\pi m^2 r^2/(qQ)$ defined in Eq. (\ref{eq:8}) and the quantity $z$ defined in
Eq. (\ref{eq:11}), which is the value of $w$ on the core surface, where $r=R$. 
Then we have the simple relation
 \begin{equation}
 v^4 w y = 1
 \label{eq:42}
 \end{equation}
between the radial variable $v = \sqrt{2M/r}$, the normalized inverse electric
field $w$, and the normalized charge $y$.  On the core surface, this becomes
 \begin{equation}
 T^4 z y = U^4 z y = V^4 z y = 1.
 \label{eq:43}
 \end{equation}

Of course, for any particular core collapse, $M$ and hence $Z$ are constants
(since we are ignoring the tiny fraction of the total energy being carried away
by the outgoing positrons and will later confirm that this fraction is indeed
small, for $M \ll 3\times 10^6 M_\odot$).  On the other hand, $Q$, $E$,
$\mathcal{N}$, $v$, $w$, and $y$ are scalar fields over the spacetime,
functions of the spacetime point, depending on both the radial and time
coordinates $(t,r)$ or $(T,X)$, or on the radial null coordinates
$(\tilde{U},\tilde{V})$ or $(U,V)$, under the assumed spherical symmetry so
that they do not depend on the angular coordinates $(\theta,\phi)$.

An intermediate example is the normalized surface inverse electric field
variable $z$, which is directly defined only on the core surface and depends
only on the single variable time variable $T=U=V=v$ on that surface where this
triple equality applies.  However, below we shall often extend the definition
of $z$ to the region outside the core as either $z(U)=w(U,U)$, the value of $w$
at the core surface (where $U=V$) at the same value of $U$ as the field point
outside (where $U<V$), or as $z(V)=w(V,V)$, the value of $w$ at the core
surface at the same value of $V$ as the field point outside.  That is, for a
field point outside the core surface with radial null coordinates $(U,V)$
(suppressing the irrelevant angular coordinates, since with the assumed
spherical symmetry, all the fields we are interested in, such as the normalized
inverse electric field $w$, are independent of them), $z(U)$ is the value of
$w$ at the point on the core surface that is to the past of the field point,
along the radial null geodesic that in the future direction of time goes
radially outward, with constant $U$, from the point on the core surface where
$z(U)=w(U,U)$ is evaluated to the field point at $(U,V)$.  Similarly, $z(V)$ is
the value of the normalized inverse electric field $w$ at the point on the core
surface that is to the future of the field point, along the other radial null
geodesic that in the future direction of time goes radially inward, with
constant $V$, from the field point $(U,V)$ to the point on the core surface
where $z(V)=w(V,V)$ is evaluated.  If the time coordinate $T=(U+V)/2$ is taken
to run vertically upward, and the radial coordinate $X=(V-U)/2$ is taken to run
horizontally to the right, so that the null coordinate $U=T-X$ increases upward
to the left and the other null coordinate $V=T+X$ increases upward to the
right, then the core surface is at the vertical line $X=0$, and the three
points $(U,U)$, $(U,V)$, and $(V,V)$ form a right triangle with the right angle
at the field point $(U,V)$ outside the core, and the hypotenuse running
vertically upward along the core surface (i.e., in the future direction of time
along the surface) from $(U,U)$ to $(V,V)$. (Here of course $U$ and $V$ are the
values they have at the field point, with $U<V$.)

Using the normalized spacetime-dependent charge $y = v^{-4} w^{-1}$ instead of
the charge in Planck units, $Q = (4\pi m^2 M^2/q)y$, the partial
differential equation (\ref{eq:27}) for the pair production becomes
 \begin{equation}
 y_{,UV} = -{Z e^{2\sigma}\over 2M^2} v^4 y^2 e^{-v^{-4}y^{-1}}
         = -{Z e^{2\sigma}\over 2M^2 v^4}{e^{-w}\over w^2}.
 \label{eq:44}
 \end{equation}
With the null coordinates $U=T-X$ and $V=T+X$ that were chosen to be equal to
$v$ on the collapsing core surface, inserting the exact and approximate values
of $e^{2\sigma}$ and of $v$ into this differential equation gives
 \begin{eqnarray}
 y_{,UV} \!&=&\! {1\over 4}(y_{,TT}-y_{,XX})
                     \nonumber \\
         \!&=&\! -{8Z(1-v^2) v^4 y^2 \over U^4(1-U) V^4(1+V)} e^{-v^{-4}y^{-1}}
	             \nonumber \\
         \!&=&\! -{8Z(1-v^2)\over U^4(1-U) V^4(1+V) v^4} {e^{-w}\over w^2}
                     \nonumber \\
         \!&\approx &\! -{128Z \over (U+V)^4} y^2 e^{-v^{-4}y^{-1}}
	             \nonumber \\
	 \!&\approx &\! -{2^{15}Z \over (U+V)^{12}} {e^{-w}\over w^2}
	              \nonumber \\
	 \!&\approx &\! -{8Z\over T^4} y^2 e^{-v^{-4}y^{-1}}
	              \nonumber \\
	 \!&\approx &\! -{8Z\over T^{12}} {e^{-w}\over w^2}.
 \label{eq:45}
 \end{eqnarray}

As with $Q$, $y=qQ/(4\pi m^2 M^2)$ satisfies the boundary conditions $y_{,U}=0$
at the core surface ($U=V$), so that there is no outward flux of positrons
there, and $y_{,V}=0$ at past null infinity ($U=-\infty$), so that there is no
inward flux of electrons there.

To calculate the time rate of change of the normalized charge $y$ at the
surface, $y_{,T}=y_{,V}$ there where $y_{,U}=0$, we can integrate $y_{,UV}dU$
radially inward along the inward null geodesic $V=\mathrm{const.}$ from
$U=0$ at past null infinity to $U=V$ at the core surface where we want to
evaluate $y_{,V}$.  However, to do this by Eq. (\ref{eq:45}), we need to know
$y(U,V)$ all the way along this null geodesic, which is determined exactly by
solving the nonlinear Eq. (\ref{eq:45}), with its boundary conditions, in the
past wedge between the core surface and this inward null geodesic.

Here I shall not attempt to determine the exact solution (which almost
certainly would not be expressible in closed form in terms of elementary
functions) or even to solve the partial differential equation (\ref{eq:45})
numerically (which will be left for an enterprising reader; I'm content to get
a citation rather than another paper for this).  Instead, I shall construct an
approximate solution, first reducing the partial differential equation
(\ref{eq:45}) to an approximate ordinary differential equation for $y(T)=y(v)$
or for $z(T)=z(v)$ on the core surface, and then finding explicit approximate
solutions and numerical solutions of this approximate ordinary differential
equation.

I shall use the fact that most of the pair production is confined to a region
close to the surface, $X \ll T^2$, where $r=2M/v^2 \approx (2M/T^2)(1+2X/T^2)$
has not risen far above its surface value $2M/T^2$, so that the electric field
has not dropped far below its surface value and so that the normalized inverse
electric field variable $w$ has not risen relatively far above its value $z$
at the surface.  (From the dominant factor $e^{-w}$ in the pair production rate,
the pair production becomes relatively negligible once $w$ increases by a few
units above its value $z$ at the surface, but since $z \gg 1$, $w$ increases by
a few units when its change is still relatively small in comparison with $w$ or
$z$.)

Outside the dominant pair-production region, there may be a significant outward
current flux $q\mathbf{n_+} = qn_+^V \partial_V \propto -y_{,U}\partial_V$ of
positrons flowing outward from the pair-production region, but the inward
current flux $-q\mathbf{n_-} = -qn_-^U \partial_U \propto +y_{,V}\partial_U$
will become negligible once one gets sufficiently far outside the dominant
pair-production region (since there is assumed to be no current coming inward
from infinity).  Therefore, outside the dominant pair-production region, $y$
will be nearly constant along the outward null geodesics, $U=\mathrm{const.}$. 
Within the pair-production region, the inward number flux 4-vector
$n_-^U\partial_U$ of electrons will be future directed, so $n_-^U$ will be
positive and hence $y_{,V}$ will be negative.  This means that as one moves
outward along a outward null geodesic $U=\mathrm{const.}$ from the core
surface, $y$ will initially decrease a bit and then rapidly settle down to near
its asymptotic value, the charge as evaluated a long way away at fixed retarded
time $U$.  (For this one should avoid going so far away that the tiny
difference from the speed of light of the velocities of the outgoing positrons
becomes significant, but that would be at an enormously large radius.  In this
paper, unless explicitly stated otherwise, I am assuming that the electrons and
positrons propagate effectively at the speed of light after their enormous
electrostatic accelerations that give each of them an increase of their energy
equal to their rest mass $m$ each time they travel a radial distance $w/(\pi m)
\ll r$.)

Therefore, a zeroth-order approximation would be to take $y$ to be constant
along the outgoing null lines, so that 
 \begin{equation}
 y(U,V) \approx y_1(U,V) \equiv y(U,U) = U^{-4} w(U,U)^{-1} = U^{-4} z(U)^{-1}.
 \label{eq:46}
 \end{equation}
 
Henceforth I shall assume that $z = z(U) = w(U,U)$, the value of $\pi E_c/E
\equiv \pi m^2/(qE)$ at the core surface at the same retarded null coordinate
$U$ as at the field point $(U,V)$ that is along an outward future-directed null
geodesic from the point $(U,U)$ on the core surface.  Since the electric field
decreases as one goes outward from the surface, whereas on the surface it
increases with time as the core contracts, $z$ is inversely proportional to the
highest value of the electric field to the causal past of the field point
$(U,V)$ and is the lowest value of $w$ to the causal past of $(U,V)$.

Now write the normalized charge $y(U,V)$ as the zeroth-order part $y_1(U,V)$
that actually depends only on $U$, plus a remainder $y_2(U,V)$ that vanishes at
the core surface $U=V$ and which will generally be smaller than $y_1(U,V)$ by a
factor of roughly $1/z$:
 \begin{equation}
 y(U,V) = y_1(U,V) + y_2(U,V) = z^{-1} U^{-4} + y_2(U,V).
 \label{eq:47}
 \end{equation}

In the region near the core surface where the dominant pair production
occurs, one can use Eqs. (\ref{eq:38}) and (\ref{eq:47}) to write
 \begin{equation}
 w(U,V) = v^{-4} y^{-1} \approx z + 4z{1-U\over U^2}X - z^2 U^4 y_2.
 \label{eq:48}
 \end{equation}
One then inserts this into Eq. (\ref{eq:45}) and changes coordinates from
$(U,V)$ to $(U,X)$ to obtain, for $X \equiv (1/2)(V-U) \ll U$,
 \begin{eqnarray}
 -{1\over 4}y_{2,XX}+{1\over 2}y_{2,XU}
	 \!&\approx &\! -{8Z\over T^{12}} {e^{-w}\over w^2}
	        \approx  -{8Z\over U^{12} z^2} e^{-w}
	             \nonumber \\
         \!&\approx &\! -{8Z e^{-z}\over U^{12} z^2} 
	      e^{-4z{1-U\over U^2}X + z^2 U^4 y_2}.
 \label{eq:49}
 \end{eqnarray}
Here I have used the fact that since $y_1$ depends only on $U$, it gives
$y_{1,UV} = 0$, and the fact that the relatively small $y_2$ term makes a
significant contribution on the right hand side only to the $e^{-w}$ term.
The effect of higher-order terms in $y_2$ in the exponential are negligible so
long as $z^3 U^8 y_2^2 \ll 1$, which turns out to be the case, except possibly
very near the black hole horizon during the final stages of the externally
visible part of the collapse.

The boundary conditions of this partial differential equation for $y_2$ are that
it vanishes along the core surface at $X=0$, and asymptotically for large $X$
one has $y_{2,X} = 0$ to reflect the fact that there is no incoming flux of
electrons at large radius, outside the pair production region.

Because $z \gg 1$ implies that the dominant pair production region occurs
near the core surface, it turns out that the $y_{2,XU}$ term is generally
negligible in comparison with the $y_{2,XX}$ term (again except possibly very
near the horizon of the black hole that forms as the end result of the
gravitational collapse).  Therefore, for each value of $U$ and $z = z(U)$, the
partial differential equation (\ref{eq:49}) becomes an ordinary differential
equation for $y_2(X)$.  Although the $y_2$ term in the exponent makes it a
nonlinear ordinary differential equation, it is readily solvable when one
chooses the exponent to be the dependent variable, with the solution (when added
to $y_1$ to get the total normalized charge $y = qQ/(4\pi m^2 M^2)$ inside the
sphere at radial coordinate $X$ and retarded time coordinate $U$) being
 \begin{equation}
 y \approx {1\over z U^4}\left\{1-{2\over z}
  \ln{\left[{1\over 2}(\sqrt{1+P}+1)
   -{1\over 2}(\sqrt{1+P}-1)e^{-4z{1-U\over U^2}X}\right]}
 \right\}.
 \label{eq:50}
 \end{equation}
Here
 \begin{equation}
 P \equiv {4Z e^{-z}\over U^4(1-U)^2 z^2} \equiv {4U(1-S)\over (1-U)^2},
 \label{eq:51}
 \end{equation}
where
 \begin{equation}
 S \equiv 1 - {Z e^{-z}\over U^5 z^2}.
 \label{eq:52}
 \end{equation}
Since $z = z(U)$ is independent of $X$, both $P$ and $S$ are functions only of
the retarded null coordinate $U$ that was chosen to be the value of the infall
speed $v = \sqrt{2M/R}$ of the core surface, at $r=R(U)$, relative to an
observer at fixed Schwarzschild radial coordinate $r$.

In order that there be no outgoing flux of positrons at the surface (since
there is no electric field inside the conducting core and hence no pair
production there), one needs that the partial derivative of $y$ with respect to
$U$ at fixed $V$ be zero at the core surface.  In terms of the coordinates
$(U,X)$, this condition at $X=0$ becomes
 \begin{equation} 0 = y_{,U} - {1\over 2}y_{,X}
  = y_{1,U} - {1\over 2}y_{1,X} + y_{2,U} - {1\over 2}y_{2,X}.
 \label{eq:53}
 \end{equation}
However, $y_1$ has no dependence on $X$, so the second term vanishes, and along
the surface at $X=0$ one has $y_2 = 0$, so the third term vanishes as well. 
Therefore, this boundary condition relates the rate of change of the normalized
charge evaluated at the surface, which is purely $y_1(U)$, to the radial
derivative of $y_2$ at the surface (where $y_2$ vanishes, but not its radial
derivative).  Then when one differentiates the relation $z = U^{-4} y^{-4}$
along the surface, one gets
 \begin{eqnarray}
 {dz\over dU} \!&=&\! -4{z\over U} - {1\over 2} z^2 U^4 y_{2,X}(U,0)
                     \nonumber \\
     \!&\approx&\! {-8 S z/U \over 1+U+\sqrt{(1+U)^2 - 4U S}}.
 \label{eq:54}
 \end{eqnarray}

Because this equation is evaluated at the core surface ($X=0$), the
$X$-dependence has dropped out, and one thus has a nonlinear first-order
ordinary differential equation for $z(U)$.  The boundary condition for this
equation is that at the infinite past (for the proper time $\tau$), when the
core surface radius is at $R=\infty$ and hence has $U = v = \sqrt{2M/R} = 0$ at
the surface, we assumed the highest possible value for $Q$, namely $Q=M$, so
then $y(U)$ has the initial value $y_0$ (at radial infinity)
 \begin{equation}
 y(0) \equiv y_0 \equiv {qQ_0\over 4\pi m^2 M^2}
      = {q\over 4\pi m^2 M_\odot \mu}
      \equiv {B\over \mu},
 \label{eq:55}
 \end{equation}
where $B \approx 42475$ was defined in Eq. (\ref{eq:4}) and $\mu$ is
$M/M_\odot$, the core mass in units of the solar mass $M_\odot$.  Thus
initially $z = U^{-4} y^{-1} \approx (\mu/B) U^{-4}$ is infinite at $U = v =
0$, and $S = 1$ there, so that $dz/dU \approx -4 z/U$ is negative infinite
initially.

The Newtonian limit of Eq. (\ref{eq:54}) is the limit in which $U \ll 1$, since
$U$ represents the inward velocity of the core surface that is at
$U=V=v=\sqrt{2M/R}$, relative to an observer that stays at fixed $r$.  In this
limit, one gets
 \begin{equation}
 {dz\over dU} \approx -4 {z\over U} S
  \equiv -4 {z\over U} \left( 1 - {Z e^{-z}\over U^5 z^2} \right).
 \label{eq:56}
 \end{equation}
Since this is an ordinary differential equation along the worldline of the core
surface, it is convenient to replace $U$ by $v=\sqrt{2M/R}$, which is more
obviously a coordinate along the core surface.  Then Eq. (\ref{eq:56}) becomes
 \begin{equation}
 z' \equiv {v\over z}{dz\over dv} \approx -4S
 \equiv -4\left(1-{Ze^{-z}\over v^5 z^2}\right).
 \label{eq:64}
 \end{equation}
If one uses $z = v^{-4}y^{-1}$, one can rewrite this explicit differential
equation for $z(v)$ as the following explicit differential equation for
$y(v)=y_1(v)$, the normalized charge at the core surface:
 \begin{equation}
 {dy\over dv} \approx -4 Z v^2 y^3 e^{-v^{-4}y^{-1}}.
 \label{eq:65}
 \end{equation}

These differential equations, Eqs. (\ref{eq:64}) and (\ref{eq:65}), are those
of the discharge process in the Newtonian limit of the core collapse.  They may
be readily derived in that limit by assuming that when the electric field
produces pairs of electrons and positrons, the electrons instantaneously
propagate to the core to reduce its charge, and the positrons instantaneously
propagate to large radii, so that the charge $Q$ is at each moment of time
independent of the radius.

Actually, to derive Eqs. (\ref{eq:64}) and (\ref{eq:65}) in the Newtonian
limit, one assumes that $z \gg 1$, which leads to a simplified approximate
expression for the integration of the pair production rate over all radii
outside the core surface at each instant of time.  Without this approximation,
but still taking the Newtonian limit of the core surface moving very slowly
compared to the speed of light, so that the electrons can propagate to the core
and the positrons can propagate very far from it in a time in which the core
radius changes only very little, one gets that the factors containing $Z$ on the
right hand sides should be multiplied by
 \begin{eqnarray}
 J(z) &=& \int_z^\infty e^{z-w} \left({z\over w}\right)^{3/2} dw
 = 2z - 2\sqrt{\pi z^3} e^z \:\mathrm{erfc} \sqrt{z}
 \nonumber \\
 &=& 1-{3\over 2z}+{3\cdot 5\over 4z^2}-{3\cdot 5\cdot 7\over 8z^3}
   +O(z^{-4}),
 \label{eq:66}
 \end{eqnarray}
a factor that is always just a bit smaller than unity (for the inevitably
large $z$).  Then Eqs. (\ref{eq:64}) and (\ref{eq:65}) become (after also
multiplying the latter equation by $v/y$)
 \begin{equation}
 z' \equiv {v\over z}{dz\over dv} 
 \approx -4\left(1-{Ze^{-z}J(z)\over v^5 z^2}\right) \equiv 4(1-e^{-W}),
 \label{eq:67}
 \end{equation}
 \begin{equation}
 y' \equiv {v\over y}{dy\over dv} 
 \approx -4 Z v^3 y^2 J(v^{-4}y^{-1}) e^{-v^{-4}y^{-1}} \equiv -4e^{-W},
 \label{eq:68}
 \end{equation}
where
 \begin{equation}
 W \equiv z+2\ln{z}-\ln{J(z)}-\ln{Z}+5\ln{v}.
 \label{eq:69}
 \end{equation}

Using the fact that $J(z)$ obeys the differential equation
 \begin{equation}
 {dJ \over dz} = \left(1+{3\over 2z}\right) - 1,
 \label{eq:70}
 \end{equation}
with the boundary condition that $J(z) \rightarrow 1$ for $z \rightarrow
\infty$, one can readily show that differentiating Eq. (\ref{eq:69}) gives
 \begin{equation}
 v{dW\over dv} = 5 - 4\left({z\over J(z)}+{1\over 2}\right)(1-e^{-W}),
 \label{eq:71}
 \end{equation}
so in this Newtonian analysis $W$ cannot decrease to 0, and $z$ simply
continues to decrease as $v$ increases.

One may now insert the correction factor $J(z)$ into the relativistic equations
above to get improved equations that work to all orders in $1/z$ when $v$ is
very small, though there will be error terms of order $v/z$ that I shall not
bother calculating.  Again replacing $U$ by its value $v$ on the core surface
where these ordinary differential equations apply, one can define
 \begin{equation}
 \tilde{S} \equiv 1 - {Z e^{-z} J(z)\over v^5 z^2} \equiv 1-e^{-W}
 \label{eq:72}
 \end{equation}
and
 \begin{equation}
 \tilde{P} \equiv P J(z) \equiv {4Z e^{-z} J(z) \over v^4(1-v)^2 z^2}
  \equiv {4v(1-\tilde{S})\over (1-v)^2}.
 \label{eq:73}
 \end{equation}
Then the relativistic versions of Eqs. (\ref{eq:67}) and (\ref{eq:68}) become
 \begin{equation}
 z' \equiv {v\over z}{dz\over dv} \approx 
  {-8 \tilde{S} \over 1+v+\sqrt{(1+v)^2 - 4v \tilde{S}}}, 
 \label{eq:74}
 \end{equation}
 \begin{equation}
 y' \equiv {v\over y}{dy\over dv} = -4-z' \approx
   {-8(1-\tilde{S})\over 1-v+\sqrt{(1+v)^2 - 4v \tilde{S}}}.
 \label{eq:75}
 \end{equation}
(It is not a typographical error for the first denominator to have $1+v$
outside the square root and for the other to have $1-v$; one can
straightforwardly check that the two equations are consistent with each other.)

The approximations used in my derivation of these equations are only valid for
$1-v \gg 1/z$, so they will break down as the core surface nears the black hole
horizon, where $v=1$.  However, the ordinary differential equations
(\ref{eq:74}) and (\ref{eq:75}) do not have any singular behavior as $v$
approaches unity, so it might be that they remain fairly good approximations
even as the core surface enters the black hole.  On the other hand, I have
definitely neglected various terms of the order of $v/z$, so near the horizon
there will almost certainly be relative errors of the order of $1/z$, which as
we shall see below is of the order of $2\%$.

\subsection{Approximate solutions of the ordinary differential equations}

Now let us find explicit approximations for the solutions of these ordinary
differential equations and compare the results to the numerical solutions.

First, let us define various quantities that will be useful in the analysis. 
Some useful constants (depending on the mass $M = M_\odot \mu$ of the
collapsing core) are
 \begin{equation}
 L \equiv \ln{Z} \equiv \ln{(A\mu^2)} 
   \equiv \ln{\left({q^2 m^2 M^2 \over \pi}\right)}
   \approx 65.69511 + 2\ln{\mu},
 \label{eq:76a}
 \end{equation}
 \begin{equation}
 y_0 \equiv {B \over \mu} \approx {42475 \over \mu},
 \label{eq:77}
 \end{equation}
 \begin{equation}
 \tilde{L} \equiv L + {5\over 4} \ln{y_0}
  \equiv \ln{A} + {5\over 4}\ln{B} + {3\over 4}\ln{\mu}
  \approx 79.01593 + 0.75\ln{\mu}.
 \label{eq:78}
 \end{equation}

Second, it is convenient to remember the definition
 \begin{equation}
 J(z) = 2z - 2\sqrt{\pi z^3} e^z \:\mathrm{erfc} \sqrt{z} 
 \label{eq:79}
 \end{equation}
and also define some new explicit functions of $z \equiv \pi E_c/E$ at the
collapsing core surface, such as
 \begin{equation}
 K(z) \equiv z + 2\ln{z} - \ln{J(z)} = z + 2\ln{z} 
 +{3\over 2z}-{21\over 8z^2}+{69\over 8z^3}+O(z^{-4}),
 \label{eq:80}
 \end{equation}
its derivative with respect to $z$,
 \begin{equation}
 K'(z) = {dK(z)\over dz} = 1 +{2\over z} 
 -{3\over 2z^2}+{21\over 4z^3}-{207\over 8z^4}+O(z^{-5}),
 \label{eq:81}
 \end{equation}
 \begin{equation}
 \hat{W}(z) \equiv K(z) - {5\over 4}\ln{z} - \tilde{L},
 \label{eq:82}
 \end{equation}
and
 \begin{equation}
 \hat{Y}(z) \equiv e^{-\hat{W}} \equiv Z y_0^{5/4} z^{-3/4} e^{-z} J(z).
 \label{eq:83}
 \end{equation}
Note that $W$ and $\hat{W}$ are two different functions of $z$, though they are
approximately the same in certain circumstances.

Third, it is convenient to define several explicit functions of $v \equiv
\sqrt{2M/R}$ (with $R$ being the core radius during its assumed free-fall
collapse from rest at radial infinity), such as
 \begin{equation}
 f(v) \equiv L - 5\ln{v} = \ln{\left({A\mu^2\over v^5}\right)},
 \label{eq:84}
 \end{equation}
 \begin{equation}
 z_1(v) \equiv y_0^{-1} v^{-4},
 \label{eq:85}
 \end{equation}
 \begin{equation}
 z_2(v) \equiv f(v)-2\ln{f(v)} + {4\ln{f(v)}-0.25+1.25v \over f(v)}
  + {4\ln^2{f(v)} -(8.5-2.5v)\ln{f(v)} \over f(v)^2},
 \label{eq:86}
 \end{equation}
 \begin{equation}
 X_1(v) \equiv \exp{[K(z1(v))-f(v)]}
  = {v^5 z_1(v)^2 e^{z_1(v)} \over Z J(z_1(v))},
 \label{eq:87a}
 \end{equation}
 \begin{equation}
 X_2(v) \equiv {5(1+v) \over 4 z_2(v)} - {5(1+2v)(3-2v) \over 16 z_2(v)^2},
 \label{eq:87b}
 \end{equation}
 \begin{equation}
 \bar{X}(v) \equiv X_1(v) + X_2(v),
 \label{eq:88}
 \end{equation}
 \begin{equation}
 g(v) \equiv f(v) + \ln{[1+\bar{X}(v)]},
 \label{eq:89}
 \end{equation}
 \begin{eqnarray}
 \bar{z}(v) \equiv g(v)\!\!\!\!&-&\!\!\!\!2\ln{g(v)}
    +g(v)^{-1}(4\ln{g(v)}-{3\over 2})
      +g(v)^{-2}(4\ln^2{g(v)}-11\ln{g(v)}+{45\over 8})
 \nonumber \\
 \!\!\!\!&+&\!\!\!\!g(v)^{-3}({16\over 3}\ln^3{g(v)}-30\ln^2{g(v)}
    +{89\over 2}\ln{g(v)}-{177\over 8}),
 \label{eq:90}
 \end{eqnarray}
 \begin{equation}
 \bar{y}(v) \equiv {1 \over v^4 \bar{z}(v)},
 \label{eq:91}
 \end{equation}
 \begin{equation}
 \bar{Q}(v) \equiv {4\pi m^2 M^2 \bar{y}(v) \over q}
  = {M_\odot \mu^2 \bar{y}(v) \over B},
 \label{eq:92}
 \end{equation}
 \begin{equation}
 \bar{E}(v) \equiv {\pi m^2 \over q \bar{z}(v)},
 \label{eq:93}
 \end{equation}
 \begin{equation}
 \bar{\mathcal{N}}(v) \equiv {m^2 \over 4\pi}{e^{-\bar{z}}\over \bar{z}^2},
 \label{eq:94}
 \end{equation}
and
 \begin{equation}
 \bar{\rho}(v) \equiv {\pi^2 e^{\pi} \over \bar{z}^2 e^{\bar{z}}},
 \label{eq:95}
 \end{equation}

Finally, it is helpful to define or recall some functions that are explicitly
given in terms of both $v$ and $z$, though of course after solving the
differential equation (\ref{eq:74}) for $z(v)$ or for $v(z)$, they can then be
expressed as functions purely of $v$ or of $z$, though there will not be
closed-form explicit exact forms for that functional dependence:
 \begin{equation}
 Y \equiv e^{-W} \equiv 1-\tilde{S} \equiv {Z e^{-z} J(z) \over v^5 z^2},
 \label{eq:96}
 \end{equation}
 \begin{equation}
 W \equiv -\ln{Y} \equiv -\ln{(1-\tilde{S})} \equiv K(z) - f(v),
 \label{eq:97}
 \end{equation}
 \begin{equation}
 X \equiv {1-Y \over Y} \equiv e^W - 1 \equiv \exp{[K(z)-f(v)]} - 1
 \equiv {v^5 z^2 e^{z} \over Z J(z)} - 1,
 \label{eq:98}
 \end{equation}
and
 \begin{equation}
 D \equiv {1\over 2}\left[1+v + \sqrt{(1-v)^2 + 4vY}\right]
  \equiv {1\over 2}\left(1+v + \sqrt{(1+v)^2 - {4vX \over 1+X}}\right),
 \label{eq:99}
 \end{equation}

Here we shall motivate the choice of $\bar{z}(v)$ as an excellent approximation
(for $M \leq 3\times 10^6 M_\odot$) to the solution of the differential
equation (\ref{eq:74}) for $z(v)$, which then gives $\bar{y}(v)$ as an
excellent approximation to the normalized charge $y(v)$, $\bar{Q}(v)$ as an
excellent approximation to the charge of the core surface when it has radius $R
= 2M/v^2$, $\bar{E}(v)$ as an excellent approximation to the electric field at
the core surface, $\bar{\mathcal{N}}(v)$ as an excellent approximation to the
pair-production rate at the core surface, and $\bar{\rho}(v)$ as an excellent
approximation to the ratio of the pair-production rate at the core surface to
the minimum value in a hypothetical dyadosphere, which has $E \geq E_c =
m^2/q$.

From the definitions above and the differential equation (\ref{eq:74}), one can
readily derive that with the independent variable being chosen to be $f(v)
\equiv L - 5\ln{v}$, $X \equiv \exp{[K(z)-f(v)]} - 1$ obeys the differential
equation
 \begin{equation}
 {dX\over df} \approx {4K'zX\over 5D}-X-1,
 \label{eq:100}
 \end{equation}
where here $z$ is to be defined implicitly as a function of $f$ and $X$ by
inverting the formula for $X$ to get
 \begin{equation}
 K(z) = f + \ln{(1+X)}
 \label{eq:101}
 \end{equation}
and then solving this for $z$.  For example, for $K(z) \gg 1$, which is always
the case here, an approximate solution for $z$ is
 \begin{eqnarray}
 z \approx K\!\!\!\!&-&\!\!\!\!2\ln{K}+K^{-1}\left(4\ln{K}-{3\over 2}\right)
 +K^{-2}\left(4\ln^2{K}-11\ln{K}+{45\over 8}\right)
 \nonumber \\
 \!\!\!\!&+&\!\!\!\!K^{-3}\left({16\over 3}\ln^3{K}-30\ln^2{K}
     +{89\over 2}\ln{K}-{177\over 8}\right),
 \label{eq:102}
 \end{eqnarray}

At $v=0$ or $f = L - 5\ln{v} = \infty$, one has that $4K'z = \infty$ and $D=1$,
so $X = \infty$ there.  One can see directly from Eq. (\ref{eq:75}) that the
normalized charge $y$ stays very near its asymptotic value $y(0) = y_0 =
B/\mu$ for sufficiently small $v$ (where the core is so large, and its electric
field is so weak, that there is negligible pair production), so there $z =
y^{-1}v^{-4} \approx y_0^{-1}v^{-4} = z_1(v)$.  Then $X_1(v) \equiv
\exp{[K(z1(v))-f(v)]}$ obeys the equation
 \begin{equation}
 {dX_1\over df} = \left({4\over 5}K'(z_1)z_1 - 1\right)X_1,
 \label{eq:103}
 \end{equation}
which is precisely the same as the Eq. (\ref{eq:100}) that $X$ obeys if $D=1$,
if $z$ is replaced by $z_1(v)$, and if the final 1 is omitted from Eq.
(\ref{eq:100}).

$D \approx 1$ when either $X \gg 1$ or $v \ll 1$ (or both), so that one may
then expand Eq. (\ref{eq:99}) for $D$ to second-order in $v$ to get
 \begin{equation}
 D \approx 1 + {v \over 1+X} + {v^2 X \over (1+X)^2}.
 \label{eq:103b}
 \end{equation}
Therefore, until $v = \sqrt{2M/R}$ gets to be near unity (its value when the
core surface collapses into the event horizon at $R=2M$), it is a fairly good
approximation to take $D \approx 1$.

The relation between $z_1(v)$ and $X_1(v)$ is
 \begin{equation}
 K(z_1(v)) = f(v) + \ln{X_1(v)},
 \label{eq:104}
 \end{equation}
which differs from Eq. (\ref{eq:101}) in not having the additive term 1 in the
argument of the logarithm, so $X \approx X_1$ implies that $|z-z_1| \ll 1$ only
if $X_1 \gg 1$.  However, until $X_1$ drops to become much smaller than unity
(e.g., for all $X_1(v) \gg z_1(v)^{-1} \ll 1$), one still has $z_1 \gg 1$, so as
far as the relative error is concerned, one still has $z \approx z_1$ in Eq.
(\ref{eq:100}).

Finally, the final 1 in Eq. (\ref{eq:100}) is negligible so long as
$z_1(v)X_1(v) \gg 1$, which for $\mu = M/M_\odot \ll B \approx 42475$ is
sufficient to imply that both $D \approx 1$ and $z_1 \gg 1$.  Therefore, under
these conditions (sufficiently small $v$), $X_1(v)$ is a good approximation for
$X(v)$.  Then one may use Eqs. (\ref{eq:101}) and (\ref{eq:102}) to get a good
approximation for $z(v)$ for these small values of $v$.

Now let us discuss what happens when we go to larger values of $v$, so that $zX$
becomes no longer large and $X_1(v)$ is no longer a good approximation to $X$. 
If $z$ and $D$ are taken to be known functions of $v$ (which of course is not
really true before one finds the solution, but one can use crude approximations
for them), then Eq. (\ref{eq:100}) is a linear equation in $X$ and hence has, as
$v$ increases and $f$ decreases, $X$ exponentially approaching a separatrix. 
Therefore, for $v$ significantly larger than its value where $zX$ becomes of
order unity, $X$ should have become very near the separatrix.  Thus we can
calculate an approximation for the separatrix in order to get an excellent
estimate for what $X$ (and hence $z$) is for large $v$ (e.g., $v=1$, where the
core surface enters the black hole horizon).

This behavior of the differential equation (\ref{eq:100}) is a consequence of
the physical fact that the pair production is a highly regulated process.  Once
pair production becomes significant to eject charge at a logarithmic rate
comparable to that of the collapsing radius $R(t)$ (which is far, far before
dyadosphere values are reached, since the astrophysical length and time scales
of the collapsing core are far greater than the Compton wavelength of the
electrons and positrons being produced), the pair production keeps discharging
the core as its radius contracts and prevents its electric field from rising
very fast.

As a first approximation, the right hand side of Eq. (\ref{eq:100}) tends to
zero, so that $X$ tends to become nearly constant.  But of course as $z$ (and,
to a lesser extent, $D$) changes, the right hand side of Eq. (\ref{eq:100})
cannot be precisely zero without $X$ changing, so one really needs to keep both
the left hand side and the right hand side slightly nonzero along the
separatrix that $X(f(v))$ approaches.

One can postulate that the separatrix can be written as a power series in
$z^{-1}$ (which stays small all the way to $v=1$), with coefficients that are
functions of $v$.  One can then show that to second order in $z^{-1}$, the
separatrix is at
 \begin{equation}
 \hat{X}(v,z) \approx {5\over 4}(1+v)z^{-1} - {5\over 16}(1+2v)(3-2v)z^{-2}.
 \label{eq:105}
 \end{equation}

One can now insert this in place of $X$ into Eq. (\ref{eq:101}) to obtain an
equation for $z(f)$, after inverting the definition $f(v) \equiv L - 5\ln{v}$
in Eq. (\ref{eq:84}) to get $v = \exp{[(L-f)/5]}$ as a function of $f$. 
Because Eq. (\ref{eq:105}) gives $\hat{X}(v,z)$ as an explicit function of both
$v$ (or of $f$) and of $z$, inserting it into Eq. (\ref{eq:101}) will give
$z$-dependence on both sides of that equation, so one cannot directly use Eq.
(\ref{eq:102}) as an approximate solution for $z(v)$ or $z(f)$.  However, one
can find that an approximate solution for this large-$v$ regime is given by the
function $z_2(v)$ defined by Eq. (\ref{eq:86}).  If we then insert this
approximation for $z$ into Eq. (\ref{eq:105}) for the approximate separatrix
given $\hat{X}(v,z)$, then this becomes the explicit function $X_2(v)$ given by
Eq. (\ref{eq:87b}).  This is then an excellent approximation to $X(v)$ at large
$v$, where the approximation valid at small $v$, $X_1(v)$, becomes much smaller
than $1/z$.

Thus we have produced approximations for both small and for large $v$, divided
by a transition region where the right hand side of Eq. (\ref{eq:100}) is of
the order of unity.  However, we would ideally like a single formula that
applies for all $v$, including the transition region.  It turns out that a
simple procedure works quite well, essentially because $v$, $z$, and $D$ change
very little across the transition region where the right hand side of Eq.
(\ref{eq:100}) rapidly drops from being much larger than unity for smaller $v$
to much smaller than unity for larger $v$.  For smaller $v$, where $X_1(v)$ is
an excellent approximation to $X(v)$, $X_1(v) \gg X_2(v)$, but for larger $v$,
where $X_2(v)$ is an excellent approximation to $X(v)$, $X_2(v) \gg X_1(v)$. 
Therefore, in each case the better approximation dominates, and one can simply
take their sum as the excellent approximation $\bar{X}(v) \equiv X_1(v) +
X_2(v)$ given by Eq. (\ref{eq:88}).

When $\bar{X}(v)$ is inserted as an approximation to $X$ in Eq. (\ref{eq:101}),
and when Eq. (\ref{eq:102}) is then used to get an explicit approximate
solution of this equation for $z(v)$, one obtains $\bar{z}(v)$ given by Eq.
(\ref{eq:90}), where the $g(v)$ used therein is defined by Eq. (\ref{eq:89}) as
what one would get for $K(z)$ from Eq. (\ref{eq:101}) with $\bar{X}(v)$ used in
place of $X$.  As a result, we have obtained a completely explicit approximation
$\bar{z}(v)$ for $z(v)$ for all $v \equiv \sqrt{2M/R}$ from $v=0$ (core surface
radius $R=\infty$) to $v=1$ (core surface radius $R=2M$, crossing the black hole
horizon).

The only part of the explicit algorithm for getting this approximate expression
$\bar{z}(v)$ that is not an elementary function is the definition of $J(z)$ in
Eq. (\ref{eq:79}) in terms of the complementary error function.  However, an
elementary function that gives an excellent approximation to $J(z)$ for the
large $z$ that occur in the problem, with error only $O(1/z^5)$ that turns out
to be numerically less than $2\times 10^{-8}$ for all values of $z$ that occur,
is
 \begin{equation}
 \bar{J}(z) =
  \left[1+\left({9\over 2}+\sqrt{6}\right)z^{-1}\right]\
   ^{^{\left(-{11\over 38}+{7\over 76}\sqrt{6}\right)}}
  \left[1+\left({9\over 2}-\sqrt{6}\right)z^{-1}\right]\
   ^{^{\left(-{11\over 38}-{7\over 76}\sqrt{6}\right)}}.
 \label{eq:106}
 \end{equation}
If this formula is used instead of $J(z)$ from Eq. (\ref{eq:79}) in Eq.
(\ref{eq:80}) for $K(z)$ and their subsequent use in Eq. (\ref{eq:87a}) for
defining $X_1(v)$ that then goes into Eq. (\ref{eq:88}) for defining
$\bar{X}(v)$, which then in turn is used in Eq. (\ref{eq:89}) to define $g(v)$,
which finally is inserted into Eq. (\ref{eq:90}) to get the final approximation
$\bar{z}(v)$ for $z(v)$, we then have a completely explicit elementary (though
admittedly slightly involved) formula for an approximation to the solution to
the relativistic ordinary differential equation (\ref{eq:74}), which itself
came from an approximate solution to the covariant partial differential
equation (\ref{eq:28}) for the pair production and discharge throughout the
spacetime region outside the surface of the collapsing core (assumed to be
freely collapsing in with a Schwarzschild metric outside, ignoring the
electrical and gravitational effects of the electric field, both which would
slow down the collapse and increase the time for discharge, thereby increasing
$z(1)$ and reducing the pair production rate).

If we would like a simpler approximate formula (though of course less
accurate), we may drop the factors of $J(z)$ (since for $z > 57.6$ it is always
within 2.5\% of unity), replace $X_2(v)$ by zero (since it is always small), and
take only the first two terms of Eq. (\ref{eq:90}).  This then gives the
approximation for $z(v)$ that I shall call $\tilde{z}(v)$,

 \begin{eqnarray}
 \tilde{z}(v) \!\!\!\!&\equiv&\!\!\!\!
  \ln{\left(A\mu^2 v^{-5} + B^{-2}\mu^2 v^{-8} e^{B^{-1}\mu v^{-4}}\right)}
 \nonumber \\
 \!\!\!\!&-&\!\!\!\! 2
 \ln{\ln{\left(A\mu^2 v^{-5} + B^{-2}\mu^2 v^{-8} e^{B^{-1}\mu v^{-4}}\right)}}.
 \label{eq:90b}
 \end{eqnarray}
 
Once we have the approximate solution $\bar{z}(v)$ for $z(v) = \pi E_c/E$,
where $E$ is the electric field strength at the surface of the freely
collapsing core when it has radius $R = 2M/v^2$, we can easily evaluate other
quantities on the core surface.  For example, an approximation to the
normalized value $y$ of the charge $Q$ is given by $\bar{y}(v) \equiv 1/[v^4
\bar{z}(v)]$ from Eq. (\ref{eq:91}), an approximation to the charge itself is
given by $\bar{Q}(v) \equiv 4\pi m^2 M^2 \bar{y}(v)/q$ from Eq. (\ref{eq:92}),
an approximation to the electric field $E$ is given by $\bar{E}(v) \equiv \pi
m^2/[q \bar{z}(v)]$ from Eq. (\ref{eq:93}), an approximation to the pair
production rate per four-volume is given by $\bar{\mathcal{N}}(v) = m^2/(4\pi
\bar{z}^2 e^{\bar{z}})$ in Eq. (\ref{eq:94}), and an approximation to the ratio
of this rate to the minimal dyadosphere rate $\mathcal{N}_c$ (that of the
critical electric field $E_c = m^2/q$) is given by $\bar{\rho}(v) \equiv
\bar{\mathcal{N}}(v)/\mathcal{N}_c = \exp{(\pi-\bar{z})}\pi^2/\bar{z}^2$ in Eq.
(\ref{eq:95}).

Of course, one can do the analogous procedure from the simpler approximate
solution $\tilde{z}(v)$ to define $\tilde{y}(v) \equiv 1/[v^4 \tilde{z}(v)]$,
$\tilde{Q}(v) \equiv 4\pi m^2 M^2 \tilde{y}(v)/q$, $\tilde{E}(v) \equiv \pi
m^2/[q \tilde{z}(v)]$, $\tilde{\mathcal{N}}(v) = m^2/(4\pi \tilde{z}^2
e^{\tilde{z}})$, and $\tilde{\rho}(v) \equiv
\tilde{\mathcal{N}}(v)/\mathcal{N}_c = \exp{(\pi-\tilde{z})}\pi^2/\tilde{z}^2$.

Now let us compare these explicit approximate expressions with the numerical
solution to the ordinary differential equation (\ref{eq:74}).  When this was
solved numerically with Maple for $1 \leq\mu \leq 3\times 10^6$ (core mass $M$
between one solar mass $M_\odot$ and three million solar masses), it was found
that the more detailed approximate solution (barred quantities) agreed
extremely well with the numerical solution over the entire range of $v$.  For
example, for $\mu=1$, it was found that the approximate solution $\bar{z}(v)$
differed from the numerical solution $z(v)$ by a relative error that was always
less than $3.2\times 10^{-5}$!  The relative error had its maximum absolute
value at the peak of a very small blip in the transition region at $v \approx
0.0236144$, where $[\bar{z}(v)-z(v)]/z(v) \approx -0.0000317$.

The relative error was very much smaller for $v$ significantly lower than the
transition region, which is not surprising, since there the pair production
rate is negligible, so that it is a very good approximation just to take the
charge to have its initial value (which was $Q=M$ at very large radii, but
ignoring the effect of the energy density of the electric field on the metric
and infall rate).  For $v$ above the transition region, the relative error was
roughly constant and positive, but with a value significantly smaller than the
magnitude of the small negative peak in the transition region.  A characteristic
value of the relative error for the larger $v$ region is given by the value at
$v=1$, when the core surface just enters the black hole horizon.

A quadratic fit of the numerical solution for $z(1)$ at the horizon ($v=1$)
versus $\mu$ for $\mu=1$ ($M=M_\odot$), $\mu=\sqrt{10}$, and $\mu=10$
($M=10M_\odot$) gave
 \begin{equation}
 z(1) \approx 57.60480 + 1.932412 \ln{\mu} + 0.0010297 \ln^2{\mu},
 \label{eq:107}
 \end{equation}
whereas the explicit formula gave the quadratic fit
 \begin{equation}
 \bar{z}(1) \approx 57.60483 + 1.932414 \ln{\mu} + 0.0010281 \ln^2{\mu},
 \label{eq:108}
 \end{equation}
so the relative error in $z$ at the horizon was about $5\times 10^{-7}$, or
about one part in two million.  Although this seems remarkably small, it is
actually not too surprising, since $X(f)$, and hence also $z(v)$, rapidly
approach the separatrix as $v$ is increased beyond the transition region where
the pair production starts to become important.

One can see that the approximate estimate given in Eq. (\ref{eq:17}) in Section
2.3, before the differential equations for the pair production and core
discharge were derived and solved, $z(1) \approx 57.58846 + 1.932873 \ln{\mu}$,
is also quite close to the numerical result, with a relative difference of only
about $-0.0003$.  Therefore, the approximate estimate made there was quite
good, though its difference from the numerical result for $z(1)$ is about 600
times the difference between the much more sophisticated estimate of
$\bar{z}(1)$ and the numerical result for $z(1)$.

The simpler approximate solution $\tilde{z}(v)$ of Eq. (\ref{eq:90b}) has a
somewhat larger error, though it always appears to be within 1\% of the correct
answer.  The maximum absolute value of the relative error appears to occur at
the black hole horizon ($v=1$) when $\mu \approx 3\,693\,360$, where the
numerical solution gives $z(1) \approx 88.29447$ but $\tilde{z}(1) \approx
87.46451$, with a relative error of about -0.940\%.  For $\mu < 3\,539\,000$,
the relative error of $\tilde{z}(v)$ always seems to have magnitude always less
than 0.5\%.  For example, for $\mu=1$ ($M=M_\odot$), the maximum absolute value
of the relative error is again at the horizon, $v=1$, as it always seems to be
for any $\mu$, and at this value of $\mu$, the relative error of $\tilde{z}(v)$
is about -0.486\%.

One might think that the simpler approximate solution is thus adequate, but
because $z$ is large and is exponentiated in the formula for the upper limit on
the pair production rate $\mathcal{N}$, even a small relative error in $z(v)$
can lead to a large relative error in $\mathcal{N}$, as we shall discuss in
more detail below after giving the numerical results for $\mathcal{N}$.  The
result is that although $\tilde{z}(v)$ is adequate for getting an estimate,
within 1\% of the correct value, of $z(v)$, and hence also of the upper limit
of the electric field $E = \pi m^2/(q z)$ at the collapsing surface of the core
of mass $M = M_\odot \mu$ when it has radius $R = 2M/v^2$, $\tilde{z}(v)$ and
the corresponding $\tilde{\mathcal{N}}(v)$ are not adequate, even within 30\%
accuracy, for giving the upper limit on the pair production rate $\mathcal{N}$.

One can now also use the numerical results for $\mu=1$, $\mu=\sqrt{10}$, and
$\mu=10$ to calculate that for a solar mass core collapsing freely inward from
rest at radial infinity with initially maximal charge $Q=M$ (but assuming the
Schwarzschild metric and infall rate), the (extremely conservative) maximal
ratio of the electric field to that the critical field of a hypothetical
dyadosphere, the value at the horizon at $R=2M$, is
 \begin{equation}
 {E(2M)\over E_c} = {\pi\over z(1)}
  \approx 0.05453699 - 0.001824768 \ln{\mu} + 0.0000540770 \ln^2{\mu}.
 \label{eq:109}
 \end{equation}
The ratio corresponds to an upper limit on the electric field itself of
 \begin{eqnarray}
 E_{\mathrm{max}}\!\!\!\!&\equiv&\!\!\!\!E(2M)
  \\
  \!\!\!\!&\approx&\!\!\!\!7.216801\! \times 10^{14}\! \:\mathrm{V/cm}
  \left[1\! -\! 0.03345927 \ln{\left({M\over M_\odot}\right)}\!
   +\! 0.000991565 \ln^2{\left({M\over M_\odot}\right)}\! \right].
   \nonumber
 \label{eq:109b}
 \end{eqnarray}
$E(2M)/E_c$ is again close to the estimate of Eq. (\ref{eq:18}), $0.054552465 -
0.001830974 \ln{\mu}$.  Thus we have confirmed that indeed the value of the
electric field of a collapsing core is always less than 5.5\% of that of a
dyadosphere, differing from that of a dyadosphere by a factor that is always
more than 18.  Of course, in reality it would be unlikely ever to get a
collapsing core with $Q$ anywhere near $M$, so this is only an extremely
conservative upper limit on the electric field.

Now we can also use the numerical result to calculate an upper limit on the
pair production rate $\mathcal{N}$, say as a ratio of that rate with the
minimum pair production rate $\mathcal{N}_c$ of a dyadosphere, with $\mu \equiv
M/M_\odot$:
 \begin{eqnarray}
 \rho(1) &\equiv& {\mathcal{N}(M)\over \mathcal{N}_c}
   = {\pi^2 e^{\pi} \over z(1)^2 e^{z(1)}}
   \nonumber \\
 &\approx& {6.61168\times 10^{-27}\over \mu^2}
    (1 + 0.0005545  \ln{\mu} - 0.00001759 \ln^2{\mu}).
 \label{eq:110}
 \end{eqnarray}
Therefore, assuming that $M=M_\odot$ or $\mu=1$ is a very conservative lower
limit on the mass of a core that can collapse into a black hole, we indeed see
that even if one can somehow start with $Q=M$ when the core is very large, and
somehow not have the charge on the core itself directly ejected by the enormous
electrostatic forces (other than the discharge by the pair production process),
the pair production rate would always be more than 26 orders of magnitude
smaller than that of a putative dyadosphere.  In particular, the minimum
dyadosphere value would be more than $1.5\times 10^{26}$ times larger.

If one wants to have an explicit formula for the pair production rate, this is
one place where it helps to have the more sophisticated estimate $\bar{z}(1)$
of Eq. (\ref{eq:90}) for $z(1)$ rather than just the crude estimate of Eq.
(\ref{eq:17}), or the crude estimate $\tilde{z}(1)$ of Eq. (\ref{eq:90b}),
since $-z$ is exponentiated in calculating the rate.  For example, for $\mu=1$,
using $z_*$ instead of $z(1)$ in Eq. (\ref{eq:110}) would have led to the
estimate of $6.67244\times 10^{-27}$, which is too large by about 1.7\%, and
using $\tilde{z}(1)$ would have led to $8.83141\times 10^{-27}$, which is too
large by about 33.6\%.  On the other hand, using $\bar{z}(1)$ instead of $z(1)$
in Eq. (\ref{eq:110}) leads to Eq. (\ref{eq:95}) and the estimate
$\bar{\rho}(1) \approx 6.61149\times 10^{-27}$ for $\mu=1$, which differs from
the numerical result by less than 0.003\%.

If, for different values of $\mu = M/M_\odot$, we wish to estimate $z(1)$ and
the maximal ratio, $\rho(1)$, of the pair production rate to that of a
dyadosphere, we can simplify the algorithm leading to Eqs. (\ref{eq:90}) and
(\ref{eq:95}) for $\bar{z}(1)$ and for $\bar{\rho}(1)$ by using the fact that
at the horizon ($v=1$) (at least for $M \leq 3\times 10^6 M_\odot$; see below),
Eq. (\ref{eq:87b}) gives $X_2(1)$ greatly dominating over $X_1(1)$ of Eq.
(\ref{eq:87a}), so that 
 \begin{equation}
 X(1) \approx \bar{X}(1) \approx X_1(1) \equiv 
  {5 \over 2 z_1(1)} - {15 \over 16 z_1(1)^2} \approx
  {5 \over 2 z(1)} - {15 \over 16 z(1)^2}.
 \label{eq:111}
 \end{equation}
Then one gets the simpler explicit approximations for $z(1)$ and for $\rho(1)$
as
 \begin{eqnarray}
 \hat{z}(1) \equiv L\!\!\!\!&-&\!\!\!\!2\ln{L}+L^{-1}\left(4\ln{L}+1\right)
   +L^{-2}\left(4\ln^2{L}-6\ln{L}-{65\over 16}\right)
   \nonumber \\
   \!\!\!\!&+&\!\!\!\!L^{-3}\left({16\over 3}\ln^3{L}
       -20\ln^2{L}-{7\over 4}\ln{L}\right)
   \nonumber \\
  &\approx& 57.60469 + 1.932364 \ln{\mu} + 0.0011044 \ln^2{\mu},
 \label{eq:112}
 \end{eqnarray}
 \begin{eqnarray}
 \hat{\rho}(1) &\equiv& \pi^2 e^{\pi} 
  \exp{\left[-L-L^{-1}-L^{-2}\left(2\ln{L}-{23\over 16}\right)
   -L^{-3}\left(4\ln^2{L}-{39\over 4}\ln{L}-1\right)\right]}
   \nonumber \\
  &=& {\pi^3 e^\pi \over \alpha} \left({\hbar c\over GMm}\right)^2
  \exp{\left[-L^{-1}-L^{-2}\left(2\ln{L}-{23\over 16}\right)
   -L^{-3}\left(4\ln^2{L}-{39\over 4}\ln{L}-1\right)\right]}
    \nonumber \\
  &\approx& 6.72439\times 10^{-27}\mu^{-2}
  \exp{\left[-L^{-1}-L^{-2}\left(2\ln{L}-{23\over 16}\right)
   -L^{-3}\left(4\ln^2{L}-{39\over 4}\ln{L}-1\right)\right]}
    \nonumber \\
  &\approx& 6.61152\times 10^{-27}\mu^{-2}
  (1+0.00055366\ln{\mu}-0.00001764\ln^2{\mu}).
 \label{eq:113}
 \end{eqnarray}

Remember from Eq. (\ref{eq:76a}) that $L \equiv \ln{(A\mu^2)} \approx 65.69511
+ 2\ln{\mu}$, with $A \equiv q^2 m^2 M_\odot^2/\pi \approx 3.39643251\times
10^{28}$ defined in Eq. (\ref{eq:4}), and with $\mu \equiv M/M_\odot$, the
collapsing core mass in units of the solar mass $M_\odot$.  Therefore, a
reasonably good approximation for $\hat{\rho}(1)$ is $\pi^2 e^\pi/(A\mu^2)
\approx 6.72439\times 10^{-27}\mu^{-2}$, which is always within about 1.7\% of
$\hat{\rho}(1)$ given above for $\mu \geq 1$.  This is the same approximation
as that given by Eq. (\ref{eq:19}) of Section 2.2, before the differential
equations were derived and solved, so it confirms the remarkable accuracy of
that approximation.  However, as we shall see below, both $\pi^2
e^\pi/(A\mu^2)$ and $\hat{\rho}(1)$ become poor approximations for the actual
numerical values of $\rho(1)$ for $M > 3.3\times 10^6 M_\odot$, where during
the collapse the maximum electric field remains too weak to give significant
discharge, so that the self-regulation of the electric field hardly sets in and
all of the approximations above that rely upon it become poor.

By comparing the coefficients of the powers of $\ln{\mu}$ in Eqs.
(\ref{eq:112}) and (\ref{eq:113}) (which were obtained from the explicit
approximations by evaluating and differentiating at $\ln{\mu}=0$) with those of
Eqs. (\ref{eq:107}) and (\ref{eq:110}) (which were obtained from the numerical
solution by the slightly different technique of making a fit at $\mu=1$,
$\mu=\sqrt{10}$, and $\mu=10$), we can get some idea of much error there is in
the polynomial expansions (in $\ln{\mu}$) of the simpler explicit
approximations of (\ref{eq:112}) and (\ref{eq:113}) for the normalized inverse
electric field strength $z \equiv \pi E_c/E$ and the normalized pair-production
rate $\rho = \mathcal{N}/\mathcal{N}_c$ when the core surface enters the black
hole event horizon at $R=2M$ or $v \equiv \sqrt{2M/R} = 1$.

For example, the relative error of the quadratic expansion of Eq.
(\ref{eq:112}) for $\hat{z}(1)$, compared with the numerical results for
$z(1)$, is only about $-2\times 10^{-6}$ at $\mu=1$.  As $\mu$ is increased, it
becomes positive and rises to a maximum of about $7 \times 10^{-4}$ at $\mu
\sim 3 \times 10^6$, and then it decreases to cross zero at $\mu \approx
3\,420\,991$ and thereafter rapidly become more negative.  It reaches $-0.01$
(1\% relative error) at $\mu \approx 3\,660\,314$, reaches $-0.1$ (10\%
relative error) at $\mu \approx 4\,119\,411$, and reaches $-0.5$ (50\% relative
error, too small by a factor of 2) at $\mu \approx 7\,515\,968$.

On the other hand, the pair production rate relative to the minimal dyadosphere
rate, $\rho(1)$, is exponentially sensitive to changes in $z(t)$, so the
relative error of the series expansion for $\hat{\rho}$ in Eq. (\ref{eq:113}),
relative to the numerical value, has much more rapid changes with $\mu$ for
$\mu > 3\times 10^6$.  This relative error starts around $-2.5\times 10^{-5}$
at $\mu=1$ and at first gradually becomes more negative, going the furthest
below zero, to about $-1.24\times 10^{-3}$, at $\mu \approx 2\,740\,000$.  Then
the relative error heads back toward zero, crossing zero at $\mu \approx
3\,134\,777$ and thereafter rapidly becoming more and more positive.  It
reaches $0.01$ (1\% relative error) at $\mu \approx 3\,292\,166$, reaches $0.1$
(10\% relative error) at $\mu \approx 3\,457\,659$, reaches $0.5$ (50\%
relative error) at $\mu \approx 3\,578\,246$, and reaches $1$ (100\% relative
error, too large by a factor of 2) at $\mu \approx 3\,627\,829$, all before the
quadratic expansion in Eq. (\ref{eq:112}) ever leads to a relative error in its
estimate of $z(1)$ of more than about 0.7\%.

The relative error of the explicit formula $\bar{\rho}(1)$ given by Eq.
(\ref{eq:95}) at $v=1$, compared with the numerical calculation of $\rho(1)$,
increases slightly more rapidly with $\mu$ for $\mu > 3 \times 10^6$, reaching
$0.01$ at $\mu \approx 3\,283\,284$, $0.1$ at $\mu \approx 3\,456\,590$, $0.5$
at $\mu \approx 3\,577\,962$ and $1$ at $\mu \approx 3\,627\,649$.

The relative errors of the explicit formulae for the barred or hatted
quantities thus stay small until $\mu \equiv M/M_\odot$ gets into the
neighborhood of $BC \approx 3\,695\,647$.  In particular, $\bar{\rho}(1)$ stays
within 1\% of the numerical answer until $\mu$ reaches $0.8884\, BC$.  The
error arises essentially because for such large core masses, even though we
have been assuming $Q=M$ initially at infinite radius (and then ignoring the
electrostatic repulsion and gravitational effects of the electric field), the
electric field never gets large enough to lead to significant discharge. 
Therefore, the discharge process never reaches the self-regulation phase in
which the solution for $z(v)$ gets near the separatrix. Then it is no longer
true that $X_2(v) \gg X_1(v)$ for $v$ near unity, so $\bar{X}(v) \equiv
X_1(v)+X_2(v)$ is no longer an excellent approximation to $X(v)$.  For such
large $\mu$, with little discharge before the collapsing core falls through the
event horizon, it would be better simply to use $X_1(v)$ as the approximation
for $X(v)$ for all $v$.

With a bit of work, one could modify the explicit formula given above to agree
well with the numerical solution of the ordinary differential equations for
larger values of $\mu$, but it would be somewhat pointless, since for such
large values of $\mu$, if the core collapse really did start with $Q=M$ at core
radius $R=\infty$, one would need to use the Reissner-Nordstrom geometry and
also take into account the effect of the electrostatic repulsion.  On the other
hand, for $\mu \leq 3\times 10^6$ or $M \leq 3\times 10^6 M_\odot$, there is
significant discharge of the core surface by the time it reaches the horizon,
so by that time the electrostatic repulsion and the deviation from the vacuum
Schwarzschild geometry become negligible, and the separatrix solution becomes
an excellent approximation to the electric field and the discharge rate.

For example, if one uses the numerical calculations (assuming a freely
infalling core surface in the Schwarzschild geometry), one finds that by the
time the core surface crosses the horizon, $\xi \equiv Q/M$ has dropped to
$\xi=0.1$ for $\mu \approx 350\,055$, to $\xi=0.5$ for $\mu \approx
1\,818\,626$, to $\xi=0.9$ for $\mu \approx 3\,319\,075$, and to $\xi=0.99$ for
$\mu \approx 3\,737\,731$.  If one takes the value of $\mu$ that gives
$\xi=0.9$, so that the core has discharged only 10\% of its charge, one finds
that $\hat{\rho}(1)$ has a relative error of only 0.0149, and $\bar{\rho}(1)$
has a relative error of only 0.0162, relative to the numerical solution. 
However, the numerical solution itself probably has large errors in this region
of high $\xi=Q/M$ at the horizon, for then the approximations of the
Schwarzschild geometry and negligible effect of the electrostatic repulsion on
the infall will be poor.  Therefore, the numerical solution itself is probably
only valid for $\mu \ll BC \approx 3.7\times 10^6$, and for this range of
$\mu$, the explicit formulae (at least for the barred and hatted quantities,
though not for the pair production rates obtained from $\tilde{z}$) are
excellent approximations to the numerical results.

Although the numerical solution would not lead to a good estimate of $\xi(1) =
Q/M$ when the core surface enters the event horizon ($v=1$) unless $\xi(1) \ll
1$ there, it does give an upper limit on the horizon value $\xi(1)$ as a
function of $\mu = M/M_\odot$.  For example, for $\mu = BC \approx
3\,695\,647$, the numerical solution gives $\xi(1) \approx 0.98513$ as the
upper bound.  Lower values of $\mu$ lead to lower upper bounds on $\xi$, so one
can say with confidence that to form a black hole with $Q > 0.99\,M$, one needs
to start with a core that has a mass $M > BCM_\odot \approx 3.7\times 10^6
M_\odot$, and probably actually with a rather higher mass (even if one could
start the core with $Q=M$ when $R \gg M$).  I would be surprised if there were
any way that an astrophysical black hole could form that is within 1\% of the
charge limit unless it has more than four million solar masses.

Actually, if one were to have a collapsing core sufficiently massive that it
could collapse into a black hole without discharging significantly, it is very
hard to imagine how the positive charge (e.g., protons) could avoid being
electrostatically ejected.  Even if nuclear forces were somehow effective in
accomplishing that Herculean feat for neutron star cores, it would seem even
much more unlikely that one could form a neutron-star-like core of very many
solar masses, so that nuclear forces on the protons could conceivably be
effective in overcoming the huge electrostatic repulsion if there were a
significant charge imbalance.

Therefore, I would actually be surprised if any black holes of astrophysical
masses ever form within our universe (or our pocket universe with our values of
the masses and charges of the electron, proton, and neutron) with values of
their charges at all near their masses.  To put it more concretely, I would
predict that no astrophysical black hole ever has a detectable change in its
geometry given by the energy in its macroscopic electric field.  In other
words, $Q/M$ would always be so far below unity that the metric of an
astrophysical Reissner-Nordstrom or Kerr-Newman black hole would be
indistinguishable from a Schwarzschild or Kerr black hole.

The most plausible way I see to try to maximize $\xi=Q/M$ would be for a
cluster of galaxies, say of mass $M_C$ and with a small excess charge $Q_C$, to
collapse to form black holes which then Hawking radiate down to $M$ comparable
to $Q_C$.

But if the electrostatic repulsion for protons from the cluster were greater
than the gravitational attraction, the excess charge in protons would most
probably escape (since protons in objects much larger than neutron stars, such
as in galactic clusters, surely cannot be all bound by nuclear forces or other
forces stronger than the gravitational binding like the ones I have been
assuming, for the sake of argument, in the calculations above of the discharge
of a hypothetical collapsing charged stellar core).  Therefore, one must have
$qQ_C < M_C m_p$.  If we are extremely generous and say that superclusters of
mass $M_C \sim 10^{16} M_\odot \sim 10^{54}$ are gravitationally bound, despite
the presently observed gravitational repulsive effects of the dark energy or
cosmological constant, then we get an upper limit of $Q_C < 10^{36}$ in Planck
units.  But this is two orders of magnitude below a solar mass, so when Hawking
radiation reduces $M$ to get near $Q_C$, pair production would radiate the
charge away to keep $Q$ always much lower than $M$.

The only possibility of getting $Q \sim M$ that I can see for a macroscopic
black hole (i.e., not one that conceivably might be produced in a high energy
collider from a small number of elementary particles) is to have the initial
value of $Q$ larger than roughly $Q_c \sim BCM_\odot \sim 10^{44}$, but this
would require an initial mass of $M > qQ_c/m_p \sim 10^{62} \sim
10^{24}M_\odot$, which is greater than the mass of all the galaxies in the
observable universe.  Such a great mass does not seem to be gravitationally
bound so that it could collapse to form a black hole in the presence of the
large-scale repulsive gravity from the dark energy or cosmological constant. 
So I would predict that, most probably, macroscopic black holes with $Q \sim M$
will never form within our pocket universe.

\section{Energy efficiency of the pair production}

We have calculated that even with idealized conditions of a collapsing stellar
core initially somehow having $Q=M$ and somehow keeping its excess protons from
being driven off by the extremely large electrostatic forces (though admittedly
smaller than the nuclear forces within a nucleus), one cannot get the electric
field to become large enough to produce pairs at a rate per four-volume within
26 orders of magnitude of the minimal dyadosphere rate.  However, we did get
astrophysically significant pair production during this idealized process, and
so one might ask what the energy efficiency of this process is, what fraction
$\epsilon$ of the mass-energy $M$ of the stellar core is converted into
outgoing positrons, to give them total energy $\epsilon M$.  Here we shall show
that unless $M \gg M_\odot$, the efficiency is very small, $\epsilon <
1.86\times 10^{-4}\sqrt{M/M_\odot} \ll 1$.

In the calculations above of the upper limits of the electric field and of the
pair production rate (which occur at the surface of the collapsing charged core
just as it enters the black hole horizon), we assumed that the initial charge
$Q_0$ of the core (when it started what we conservatively assumed was free fall
from radial infinity) was the maximum possible, $Q_0 = M$.  Although this
indeed would give the maximum possible electric field and pair production rate,
those quantities at the black hole horizon would be nearly independent of the
initial charge $Q_0$, so long as $Q_0$ was large enough to lead to significant
pair production and hence to the self-regulation of the electric field during
the core collapse.

We found above that for $Q_0 = M$, the explicit approximation for the
normalized pair production rate, $\bar{\rho}(1)$, stays within 1\% of the
numerically calculated value for $\mu < 0.8884\,BC \approx (8/9)BC$, indicating
that then the self-regulation is sufficient to keep the explicit approximation
accurate to within about 1\% error.  If we had allowed $Q_0$ to be less than
$M$, the analysis would have had the initial value $y_0$ of the normalized core
charge $y = qQ/(4\pi m^2 M^2) = \hbar c q Q/(16\pi^2\epsilon_0 G^2 m^2 M^2)$
(where $q$ and $m$ are the charge and mass of a positron, and $M$ is the mass
of the core) at core radius $R=\infty$ or $v \equiv \sqrt{2M/R} = 0$ be not
$y_0 = B/\mu = BM_\odot/M$ but rather be $y_0 = BM_\odot Q_0/M^2$.  That is,
one would just replace $B/\mu = q/(4\pi m^2 M)$ with $(B/\mu)(Q_0/M) =
qQ_0/(4\pi m^2 M^2)$ in the formulas above.  Then the condition $B/\mu >
9/(8C)$ would become $Q_0/M > 9M/(8BCM_\odot) = [9/(8BC)]\mu \approx 3\times
10^{-7} \mu$.  For values of the initial charge-to-mass ratio of the collapsing
core greater than this, there would be sufficient pair production, discharge,
and self-regulation that the barred approximations above would be fairly good.

However, the energy emitted into positrons which escape (the electrons going
into the core or else into the black hole and hence not escaping) does depend
significantly upon the initial charge $Q_0$ (since of course if $Q_0$ is lower,
there is less charge to be emitted), so for this section I shall include that
dependence.  It is helpful in doing this to define the quantity
 \begin{equation}
 \xi = \xi(U,V)
 \equiv {Q \over M} = {4\pi m^2 M y \over q} = {\mu y \over B},
 \label{eq:114}
 \end{equation}
which is the charge-to-mass ratio at any point in the spacetime outside the
collapsing core, say labeled by the radial null coordinates $(U,V)$, since the
assumed spherical symmetry means that there is no dependence on the angular
coordinates $(\theta,\phi)$.

The initial value of this charge-to-mass ratio $\xi$, at past null and timelike
infinity outside the core, will be denoted by the constant value $\xi_0$ (with
maximum value 1, as it was assumed to be in the previous sections), the value
on and outside the collapsing core surface when it begins its hypothetical free
fall from radial infinity in the assumed Schwarzschild metric (ignoring the
electrostatic and gravitational effects of the electric field, as I have always
been doing to get overly conservative upper limits on the electric field and
pair production rate):
 \begin{equation}
 \xi_0 \equiv \xi(-\infty,V) \equiv {Q_0 \over M} = {\mu y_0 \over B}.
 \label{eq:115}
 \end{equation}

If one uses ingoing and outgoing radial null coordinates $U$ and $V$
respectively that both take the value $v=\sqrt{2M/R}$ at the surface of the
freely collapsing core when it has Schwarzschild radius $R$ (the circumference
of the core divided by $2\pi$), then the value of $\xi$ at the core surface as
a function of $v$ may be denoted as $\xi_s$,
 \begin{equation}
 \xi_s = \xi_s(v) \equiv \xi(v,v) = {\mu \over B v^4 z(v)}.
 \label{eq:116}
 \end{equation}
It is also convenient to define $\xi_\infty(v)$ as the value of the
charge-to-mass ratio $\xi$ at radial infinity out along the outward radial null
geodesic from the core surface at $U = v = \sqrt{2M/R}$:
 \begin{equation}
 \xi_\infty = \xi_\infty(v) \equiv \xi(v,\infty) = {Q_\infty\over M}.
 \label{eq:117}
 \end{equation}

If a positron is created very near the core surface, it will initially
experience the electric field of the charge $Q_s(v) = M\xi_s(v)$, which at the
core surface will be
 \begin{equation}
 E_s(v) = {Q_s(v) \over R^2(v)} = {Q_s(v) v^2 \over 4M^2}
  = {\xi_s(v) v^2 \over 4M}.
 \label{eq:118}
 \end{equation}
Then as the positron moves out very nearly along the radial null geodesic with
$U=v = \mathrm{const.}$ and $V$ increasing from $V=v$ at the surface to
$V=\infty$ at radial infinity, and as the positron passes the incoming
electrons that are created at greater radii and propagate inward to smaller
radii, the charge the positron experiences as being at smaller radii will
decrease to the asymptotic value $Q_\infty(v) = M \xi_\infty(v)$, and the
electric field will approach $M \xi_\infty(v)/r^2(v,V)$, which of course
becomes smaller and smaller as $r(U,V) = r(v,V)$ becomes larger and larger
along the outward path of the positron.

In all of this, we make the assumption, to be verified below for $M \ll
BCM_\odot$, that only a tiny fraction of the initial total energy of the core
is converted into outgoing positrons, so that the mass $M$ that appears in the
metric is very nearly a constant everywhere over the spacetime outside the
surface of the collapsing core, giving an approximately Schwarzschild metric.

In this approximately Schwarzschild metric (\ref{eq:20}), as written in terms of
radial null coordinates $U$ and $V$, if a positron is produced at some
$(U,V)$ and then is accelerated to very high gamma factors essentially along the
outward null line $U = \mathrm{const.}$, then the kinetic energy gained during
this acceleration by the electric field $Q(U,V)/r^2(U,V)$ is
 \begin{equation}
 \mathcal{E}(U,V) = \int_R^\infty {qQ dr \over r^2}
 = \int_V^\infty {qQ(U,V')e^{2\sigma(U,V')}dV'\over 2 r^2(U,V')},
 \label{eq:119}
 \end{equation}
where the integral is taken along the outward radial null line $U =
\mathrm{const.}$ along which the positron approximately travels from its
creation point at $(U,V)$ to radial infinity at $(U,\infty)$.  Almost all of
the positrons are created sufficiently deep in the electric field that their
final kinetic energies, at radial infinity, far exceed their rest mass
energies, so I shall ignore the latter.  Then one may multiply the positron
production rate by the energy gained by each positron and integrate over all of
the exterior region where the production is occurring to get the total energy
emitted during the core collapse as
 \begin{equation}
 \epsilon M = -{1\over q}\int \mathcal{E}(U,V) Q_{,UV} dU dV.
 \label{eq:120}
 \end{equation}
(The negative sign comes from the fact that $Q$ is decreasing during the
discharge, so $Q_{,UV} < 0$.)

In principle one could insert the approximation $\bar{z}(v)$ from Eq.
(\ref{eq:90}) into Eq. (\ref{eq:50}) to get an explicit approximation for the
distribution of charge $Q(U,V)$ in the region outside the collapsing core where
the dominant pair-production occurs, and then insert these results into Eqs.
(\ref{eq:119}) and (\ref{eq:120}) to do numerical integrations over $(U,V)$ to
find the total energy $\epsilon M$ emitted in positrons.  For even more
precision, one could directly solve the pair-production partial differential
Eq. (\ref{eq:27}) numerically and to determine the integrands of Eqs.
(\ref{eq:119}) and (\ref{eq:120}), but this shall be left as an exercise for
the reader.  However, here I shall resort to approximations, which, at the cost
of some precision, have the advantage of giving explicit results.

Just as the derivation of the approximate ordinary differential Eq.
(\ref{eq:74}) for the discharge of the core from the partial differential Eq.
(\ref{eq:27}) relied upon the fact that $z^{-1} \ll 1$, so I shall use this
same fact here to make an approximation valid only to zeroth and first order in
$z^{-1}$.  This fact also implies that, for $v$ not too close to unity, the
almost all of the pair production occurs close to the collapsing core surface,
near $r=R$, and this implies that $\xi(U,V) = Q(U,V)/M$ is nearly constant
along the outgoing radial null lines, equalling both $\xi_s(v)$ and
$\xi_\infty(v)$.

An exception to this is for $1-v \ll 1$, when the core surface is just about to
enter the black hole horizon with finite nonzero charge-to-mass ratio
$\xi(1,1)$, but when the pair production outside the black hole reduces the
charge-to-mass ratio at radial infinity, $\xi_\infty(v) = \xi(v,\infty)$, to
zero in the limit that $U=v \rightarrow 1$, which is the limit of $\tilde{U} =
t - r_* \rightarrow \infty$ at future null infinity, assuming for this argument
that the black hole lasts forever and does not evaporate away by Hawking
radiation.  That is, for a positively charged black hole that lasts forever and
does not have any incoming charge other than the electrons produced by its
electric field, eventually the charge will be seen to decay to approach
arbitrarily close to zero, because of the continuing pair production outside
the black hole and resulting flow of positrons to radial infinity.

(Strictly speaking, even this is not quite right, since the outgoing positrons
will not have quite the speed of light and hence will actually go to future
timelike infinity rather than to future null infinity, so that along future
null infinity itself, the charge will always remain at its initial value $Q_0$
that it has along past null infinity and spacelike infinity.  However, at any
large but finite $r$, as a function of the Schwarzschild time coordinate $t$,
which is proper time for a static observer at radial infinity, the charge will
asymptotically decrease to zero as the external electric field of the
initially-charged black hole produces charged pairs, with the electrons falling
down the hole to neutralize it and the positrons carrying off the charge to
radial infinity, even though actually at future timelike infinity rather than
strictly being at future null infinity.  Since under the conditions that the
charge self-regulation is significant, the positrons are accelerated to
enormous gamma factors by the electric field, for this paper I am employing the
approximation that the positrons go out at the speed of light and so am
ignoring the distinction between their outgoing worldlines and outgoing null
geodesics.)

For the formulas below, I shall assume that $\xi_0 > [9/(8BC)]\mu \approx
3\times 10^{-7} \mu$, so that the self-regulation of the electric field is
significant even before the collapsing charged core gets near the black hole
horizon, so that most of the decrease of $\xi_\infty(v)$ from $\xi_0$ at $v=0$
to 0 at $v=1$ occurs for $v$ not too close to 0.  Then the self-regulation of
the charge by the pair production process gives $\xi_\infty(v) \ll \xi_0$
before $v$ gets so close to unity that the charge-to-mass ratio at radial
infinity along the outgoing nearly-null positron worldlines, $\xi_\infty(v)$,
drops significantly below the charge-to-mass ratio $\xi_s(v)$ at the core
surface at the start of these outgoing positron worldlines.

In the limit that this approximation is exact, that is, to zeroth order in
$z^{-1}$, one gets that $\mathcal{E}(U,V) = qQ_s/R = {1\over 2}q\xi_s(v) v^2$,
This would give
 \begin{equation}
 \epsilon = \int_0^M {Q dQ \over M R} = {1 \over 2} \int_0^{\xi_0} v^2 \xi d\xi.
 \label{eq:121}
 \end{equation}
 
However, the charge $Q$ and the charge-to-mass ratio $\xi$ both actually
decrease a bit as one moves outward along each outward null geodesic
$U=v=\mathrm{const.}$.  If one uses Eq. (\ref{eq:50}) (with $P$ replaced by
$\tilde{P}$ that is the improved value for $v \ll 1$, though I shall remind the
reader that I do not know the improved value for $v \sim 1$ and so there am
likely to have errors of the order of $v/z$) for the charge distribution
outside the collapsing charged core surface, one gets that
 \begin{eqnarray}
 \xi(v,V) &\approx& {\mu\over B v^4 z(v)}-{2\mu\over B v^4 z(v)^2}
  \ln{\left[{1\over 2}(\sqrt{1+\tilde{P}(v)}+1)\right]} 
  \nonumber \\
   &-&{2\mu\over B v^4 z(v)^2} 
   \ln{\left[1-{\tilde{P}(v)\over(\sqrt{1+\tilde{P}(v)}+1)^2}
   e^{-4z(v){1-v\over v^2}X(v,V)}\right]},
 \label{eq:122}
 \end{eqnarray}
where here $X \equiv (V-U)/2 = (V-v)/2 \approx M(r-R)/R^2$ is the spatial radial
coordinate $X$ given by Eq. (\ref{eq:37}), not the $X \equiv (1-Y)/Y$ of Eqs.
(\ref{eq:98}) and onward, and where Eq. (\ref{eq:96}) implies that the
$\tilde{P}$ defined in Eq. (\ref{eq:73}) may also be written in terms of $v$
and of $Y(v) \equiv e^{-W} = Ze^{-z}J(z)/(v^5 z^2)$, defined in Eq.
(\ref{eq:96}), as
 \begin{equation}
 \tilde{P} = {4vY\over(1-v)^2}.
 \label{eq:123}
 \end{equation}

One can see from this that
 \begin{equation}
 \xi_\infty(v) \approx {\mu \over B v^4 z(v)}\left\{1-{2\over z(v)}
  \ln{\left[{1\over 2}(\sqrt{1+\tilde{P}(v)}+1)\right]} \right\}
 \label{eq:124}
 \end{equation}
is thus a bit less than $\xi_0(v)$, by an amount that is $O(z^{-1})$ multiplied
by a logarithm.  When $\xi_0 > [9/(8BC)]\mu \approx 3\times 10^{-7} \mu$, then
for sufficiently large $v$, $\xi_s(v)$ will decrease significantly below
$\xi_0$, and in this large $v$ region, the electric field self-regulation will
lead to $Y \approx 1$ and hence to
 \begin{equation}
 \xi_\infty(v) \approx {\mu \over B v^4 z(v)}\left(1-{2\over z(v)}
  \ln{1\over 1-v} \right).
 \label{eq:125}
 \end{equation}

This equation would imply that $\xi_\infty(v)$ would become negative for $v >
1-e^{-z(v)/2}$, which is for $v$ very near unity (very near the point where the
collapsing charged core surface enters the black hole horizon at $v=1$), but
actually the approximation is breaking down for $v$ this close to unity.  As
discussed above, one would instead expect that $\xi_\infty(v)$ should
asymptotically approach 0 as $v \rightarrow 1$, as then there is an infinite
amount of time in the region external to the black hole for the pair production
to carry away all of the charge (under the assumption that the black hole
persists forever and does not evaporate by Hawking radiation, other than the
charge emission that is of course included in the full Hawking emission
formula).  Therefore, we shall take Eqs. (\ref{eq:122})-(\ref{eq:125}) to be
approximately valid only for $-\ln{(1-v)} \ll z(v)/2$.

If we define
 \begin{equation}
 a(v) \equiv {\tilde{P}(v)\over\left(\sqrt{1+\tilde{P}(v)}+1\right)^2}
      \approx v,
 \label{eq:126}
 \end{equation}
where the $\approx$ applies only for the large-$v$ region where $\xi_s(v)$
drops significantly and $Y \approx 1$ (the region where the pair production is
significant), then we can calculate that for fixed $U=v$, the mean value of the
spacelike radial coordinate $X(U,V)=X(v,U)$ at which positrons are created
along the outward radial null geodesic is
 \begin{equation}
 \langle X \rangle \approx {v^2\over 4(1-v)z(v)\ln{[1/(1-a)]}}
   \sum_{n=1}^\infty {a^n\over n^2}.
 \label{eq:127}
 \end{equation}
Then $r \approx R(1+RX/M) = R(1+2X/v^2)$ gives
 \begin{eqnarray}
 \langle {r\over R} \rangle &\approx& 1 + F(v) \equiv
 1 + {1\over 2(1-v)z(v)\ln{[1/(1-a)]}}\sum_{n=1}^\infty {a^n\over n^2}
  \nonumber \\
 &\approx&1+{1\over 2(1-v)z(v)\ln{[1/(1-v)]}}\sum_{n=1}^\infty {v^n\over n^2},
 \label{eq:128}
 \end{eqnarray}
where again the last $\approx$ applies only for the large-$v$ region where
$\xi_s(v)$ drops significantly and $Y \approx 1$, so that $a(v) \approx v$. For
$v \ll 1$ and yet still within the region where $\xi_s(v)$ has dropped
significantly below $\xi_0$, one gets $\langle r-R \rangle \approx R/(2z)$,
just as was found in Section 2.3 for getting the approximate estimate of the
discharge rate before solving any differential equations.  The main correction
for somewhat larger $v$ is to divide this result by $1-v$, making the mean
radius of the positron production somewhat larger.  This approximation would
imply that the mean radius would go to infinity as $v \rightarrow 1$, but this
is where the approximation is breaking down.  However, for $\xi_0 >
[9/(8BC)]\mu \approx 3\times 10^{-7} \mu$ so that significant self-regulation
occurs, most of the pair production occurs for $v$ sufficiently below unity
that the approximation does remain valid, so for calculating the efficiency
$\epsilon$ in such cases, one does not need to worry about the breakdown of the
approximation for $1-v \ll 1$.

Now when $1-v \gg 1/(2z) \sim 0.01$, an estimate for the average energy (for
given $v$) gained by a positron in being electrostatically accelerated out to
radial infinity that is better than the zeroth-order (in $z^{-1}$) estimate
$qQ_s/R = qM\xi_s/R$ is the first-order estimate
 \begin{equation}
 \langle\mathcal{E}(U,V)\rangle \approx {qM\xi_\infty\over \langle r \rangle}
 \approx {q v^2 \xi_\infty(v)\over 1+F(v)},
 \label{eq:129}
 \end{equation}
because most of the work done by the electric field in accelerating the
positron from its average creation position at $r=\langle r \rangle \approx
R(1+F)$ to $r=\infty$ occurs for $r-R \gg \langle r-R \rangle$ where $Q(U,V)
\approx Q_\infty(v) = M\xi_\infty(v)$.

One can now insert this expression into Eq. (\ref{eq:120}) to get an
approximate one-dimensional integral for the energy efficiency of the
pair-production process (depending on the core mass $M=M\odot\mu$ and on its
initial charge $Q_0=M\xi_0=M\odot\mu\xi_0$):
 \begin{eqnarray}
 \epsilon = \epsilon(\mu,\xi_0)
  &\approx& -{1\over q}\int_0^1 \langle\mathcal{E}(U,V)\rangle
  {d\xi_\infty(v)\over dv} dv
  \approx \int_0^{\xi_0} {M\over \langle r \rangle} \xi_\infty d\xi_\infty
  \nonumber \\
  &\approx& -{1\over 2}\int_0^1 {v^2\xi_\infty(v) \over 1+F(v)}
  {d\xi_\infty(v)\over dv} dv.
 \label{eq:130}
 \end{eqnarray}
It was this integral that was evaluated numerically from the numerical
solution of Eq. (\ref{eq:74}), using the approximate expressions for
$\xi_\infty(v)$ given by Eq. (\ref{eq:124}) and for $1+F(v)$ given by Eq.
(\ref{eq:128}).

However, before Eq. (\ref{eq:130}) was integrated numerically, an explicit
approximate solution was found.  For this it appeared too complicated to use the
approximation $\bar{z}$ of Eq. (\ref{eq:90}), so I used instead an alternate
version of $\tilde{z}(v)$ that was given by Eq. (\ref{eq:90b}).  This alternate
simple approximation is that $z(v)$ is given approximately by $\check{z}(v)$
that is the solution to the equation
 \begin{equation}
 \check{z}^2 e^{\check{z}} 
 = A\mu^2 v^{-5} 
   + B^{-2}\mu^2 \xi_0^{-2} v^{-8} e^{B^{-1}\mu \xi_0^{-1} v^{-4}}.
 \label{eq:131}
 \end{equation}
Here, unlike in Eq. (\ref{eq:90b}), I have allowed the possibility that the
initial charge-to-mass ratio $\xi_0$ is less than unity by replacing $B/\mu$
there by $B\xi_0/\mu = qQ/(4\pi m^2 M^2)$.

In this expression, the second term on the right dominates when $v$ is
sufficiently small that the pair production is small.  Indeed, if one dropped
the first term as negligible compared to the second, this would give $\check{z}
\approx B^{-1}\mu \xi_0^{-1} v^{-4}$, so then the normalized charge would be $y
= v^{-4} z^{-1} \approx v^{-4} \check{z}^{-1} \approx B\xi_0/\mu = y_0 =
\mathrm{const.}$.  Therefore, this regime would give $\xi_s(v) \approx
\mathrm{const.}$ and hence a negligible contribution to the integral
(\ref{eq:130}) for the efficiency $\epsilon$.  That is, the integral for the
efficiency is dominated by the values of $v$ for which the first term on the
right hand side of Eq. (\ref{eq:131}) dominates, the region of significant
charge self-regulation.

If we define
 \begin{eqnarray}
 \hat{L}_0 &\equiv& \tilde{L} + {5\over 4} \ln{\xi_0}
  \equiv \ln{A} + {5\over 4}\ln{B} + {3\over 4}\ln{\mu} + {5\over 4} \ln{\xi_0}
  \nonumber \\
  &\equiv& -{1\over 4}\ln{(2^{10}\pi^9 m^2 M^2 q^{-13} Q_0^{-5})}
  \approx 79.01593 + 0.75\ln{\mu} + 1.25\ln{\xi_0},
 \label{eq:132}
 \end{eqnarray}
then the two terms on the right hand side of Eq. (\ref{eq:131}) are equal at
the transition value of $v$ that I shall call $v_t = v_t(\mu,\xi_0)$ and which
is given by the following equation in terms of the related quantity $z_t =
z_t(\mu,\xi_0)$ that is defined by the next equation after that:
 \begin{equation}
 v_t = \left({\mu\over B \xi_0 z_t}\right)^{1/4},
 \label{eq:133}
 \end{equation}
 \begin{equation}
 z_t + {3\over 4}\ln{z_t} = \hat{L}_0.
 \label{eq:134}
 \end{equation}

Under the condition that significant self-regulation of the charge occurs, that
is, for $\xi_0 > [9/(8BC)]\mu \approx 3\times 10^{-7} \mu$, a sufficiently good
approximation to the solution of Eq. (\ref{eq:134}) is
 \begin{equation}
 z_t \approx \hat{L}_0 - {3\over 4}\ln{\hat{L}_0}
  + {9\over 16}\hat{L}_0^{-1}\ln{\hat{L}_0}
  + {27\over 128}\hat{L}_0^{-2}\ln{\hat{L}_0}(\ln{\hat{L}_0}-2),
 \label{eq:135}
 \end{equation}
which one can then insert into Eq. (\ref{eq:133}) to get $v_t$.  Two rougher
but more self-contained expressions for $z_t$ are
 \begin{eqnarray}
 z_t \sim &-&{1\over 4}\ln{(2^{10}\pi^9 m^2 M^2 q^{-13} Q_0^{-5})}
 \nonumber \\
 &-&{3\over 4}
 \ln{\left[-{1\over 4}\ln{(2^{10}\pi^9m^2M^2q^{-13}Q_0^{-5})}\right]}
 \nonumber \\
 &\sim& 75.77 + 0.7426\ln{\mu} + 1.238\ln{\xi_0}.
 \label{eq:136}
 \end{eqnarray}

One can see that when $v=v_t$, both terms of the right hand side of Eq.
(\ref{eq:131}) are equal to $z_t^2 e^{z_t}$, so one gets
 \begin{equation}
 z(v_t) + 2\ln{z(v_t)} = z_t + 2\ln{z_t} + \ln{2}.
 \label{eq:137}
 \end{equation}
Therefore, the transition value of $z$ defined in this way is just a bit larger
than $z_t$, so $z_t$ is a fairly good estimate of the value of $z$ where the
self-regulation of charge starts to become important as the surface of the
charged core freely collapses inward from the idealized initial conditions of
being at rest at radial infinity and then follows radial geodesics inward in
the assumed exterior Schwarzschild metric of mass $M$.

For $\mu=1$ ($M=M_\odot$, one solar mass) and $\xi_0=1$ ($Q_0=M$, initial
charge equal to the mass so that then one would have the extreme
Reissner-Nordstrom exterior metric, though the infall is always taken to occur
in the Schwarzschild metric with no effect from the electrostatic forces), one
gets from Eqs. (\ref{eq:135}) and (\ref{eq:133}) that $z_t(\mu,\xi_0)$,
$v_t(\mu,\xi_0)$, and $R_t(\mu,\xi_0)$ take the values

 \begin{equation}
 z_{t11} \equiv z_t(1,1) \approx 75.77015,
 \label{eq:138}
 \end{equation}
 \begin{equation}
 v_t \equiv v_t(1,1) \approx 0.02360985,
 \label{eq:139}
 \end{equation}
 \begin{equation}
 {R_{t11}\over 2M} \equiv {R_t(1,1)\over M_\odot}
  = {1\over v_{t11}^2} \approx 1793.964.
 \label{eq:140}
 \end{equation}

In fact, for general $\mu = M/M_\odot$ and $\xi_0 = Q_0/M$, the critical radius
where the self-regulation starts becoming significant is
 \begin{equation}
 R_t = 2M/v_t^2 \approx 5298.024 \sqrt{\mu\xi_0 z_t/75.77015}\;\mathrm{km}
 \sim 5298\, \mu^{0.5049} \xi_0^{0.5082}\; \mathrm{km}.
 \label{eq:141}
 \end{equation}

Thus if one starts with a hypothetical charged core collapsing freely from a
very large radius into a black hole with initial charge $Q_0 \sim M$, the
self-regulation of the charge will start to become important when the the core
gets to a radius of the order of the radius of the earth, and that is when there
will start to be significant pairs being produced.  The proper time left during
the free-fall collapse after the core surface crosses this radius is then
 \begin{equation}
 \Delta\tau = {4\over 3}M/v_t^3 = {4\over 3}M_\odot(B^3\mu\xi_0^3 z_t^3)^{1/4}
 \sim 0.499\, \mu^{0.257} \xi_0^{0.762}\, \mathrm{seconds}.
 \label{eq:142}
 \end{equation}
Therefore, if $Q_0 \sim M$, the burst of positrons that will be emitted as the
core collapses from $R=R_t$ to $R=2M$ lasts a time that is of the order of one
second.

For $v > v_t$, where the pair production is significant, the first term on the
right hand side of Eq. (\ref{eq:131}) dominates, and one gets $z^2 e^z \approx
A\mu^2/v^5$.  To evaluate the integral in Eq. (\ref{eq:130}), it is simplest
to express $z$ as a function of the surface charge-to-mass ratio $\xi_s$, which
leads to the equation
 \begin{equation}
 z + {3\over 4}\ln{z} \approx \hat{L}
  \equiv \hat{L}_0 + {5\over 4}\ln{\xi_s\over \xi_0},
 \label{eq:143}
 \end{equation}
with the approximate solution
 \begin{eqnarray}
 z &\approx& \hat{L} - {3\over 4}\ln{\hat{L}}
  + {9\over 16}\hat{L}^{-1}\ln{\hat{L}}
  + {27\over 128}\hat{L}^{-2}\ln{\hat{L}}(\ln{\hat{L}}-2)
  \nonumber \\
  &\sim& z_t \left(1 + {5\over 4z_t+3}\ln{\xi_s\over \xi_0} \right).
 \label{eq:144}
 \end{eqnarray}

Then when one  evaluates the integral in Eq. (\ref{eq:130}) to first order in
$z_t^{-1}$, one gets that the efficiency of the conversion of the core mass $M$
into outgoing positrons of energy $\epsilon M$ is, from the explicit
approximation,
 \begin{eqnarray}
 \epsilon(\mu,\xi_0)
  &=& {1\over 3}\xi_0^2 v_t^2
     \left[1 - {1\over 12z_t} - {6v_t\over 5z_t} + O(z_t^{-2})\right] 
 \nonumber \\
  &\approx& \bar{\epsilon}(\mu,\xi_0) \equiv {1\over 3}\xi_0^2 v_t^2
   \left(1 - {1\over 12z_t} - {6v_t\over 5z_t}\right)
 \nonumber \\
  &\approx& 0.000185534\, \mu^{1/2} \xi_0^{3/4}
   \left(1 - 0.004901\ln{\mu} - 0.008168\ln{\xi_0}\right)
 \nonumber \\
  &\approx& 0.0001855\, \mu^{0.4951} \xi_0^{0.7418}.
 \label{eq:145}
 \end{eqnarray}

By comparison, a numerical evaluation of the integral in Eq. (\ref{eq:130})
from the numerical solution of Eq. (\ref{eq:74}) for $\xi_0=1$ and for $\mu=1$
and for $\mu=10$ and then fit linearly to $\mu^{-1/2}\epsilon(\mu,1)$ versus
$\ln{\mu}$ gave
 \begin{equation}
 \epsilon(\mu,1) \approx 0.000185467\, \mu^{1/2} (1 - 0.004980 \ln{\mu}).
 \label{eq:146}
 \end{equation}

This shows that over a range of $\mu$ of the order of unity, the explicit
approximate expression seems to agree to about 4 decimal places with the
numerical calculation, so it is apparently quite accurate.  However, I have
noted above that even my numerical calculation is based upon certain
approximations to reduce the partial differential equation (\ref{eq:27}) to the
ordinary differential equation (\ref{eq:74}), and I would expect that the
relative error is of the order of $v/z$.  For $\mu=1$ and $\xi_0=1$, most of
the positrons are produced around $v \sim v_{t11} \sim 0.024$ and $z \sim
z_{t11} \sim 76$, giving $v_{t11}/z{t11} \approx 3\times 10^{-4}$ as a crude
estimate of the relative error.  Therefore, I might guess that the results for
the energy efficiency above are accurate to about 3-4 decimal places and can be
written to roughly this precision as
 \begin{equation}
 \epsilon \approx 0.0001855 \left({M\over M_\odot}\right)^{0.495}
  \left({Q_0\over M}\right)^{0.742}.
 \label{eq:147}
 \end{equation}

To show the main dependence of the efficiency upon the mass $m$ and charge $-q$
of the electron and upon the mass $M$ and initial charge $Q_0$ of the charged
core that collapses freely from rest at infinity in the assumed external
Schwarzschild metric, let us take the first term on the right hand side of Eq.
(\ref{eq:135}) or Eq. (\ref{eq:136}) as the crude approximation $\hat{L}_0$ for
$z_t$ and then insert this into Eq. (\ref{eq:145}), with the $O(z_t^{-1})$
corrections omitted, to get, with conventional units restored,
 \begin{equation}
 \epsilon \sim {m\over 3M}\sqrt{Q_0^3\over \hbar c \epsilon_0 q
 \ln{[(2^{28}\pi^{18}\hbar^7c^7\epsilon_0^9G^2m^2q^{-13}M^2Q_0^{-5})^{-1/4}]}}.
 \label{eq:148}
 \end{equation}
For $M=M_\odot \approx 1.98844\times 10^{30}$ kg and $Q_0 =
\sqrt{4\pi\epsilon_0 G} M_\odot \approx 1.71353\times 10^{20}$ coulomb, this
crude estimate gives $\epsilon \sim 0.000182$, within 2\% of the efficiency
0.0001855 obtained numerically for a solar-mass core with maximum initial
charge, $\xi_0 = Q_0/M = 1$ in Planck units.

Another way to express the efficiency is to note that Eq. (\ref{eq:145}) may be
written as
 \begin{eqnarray}
 \epsilon(\mu,\xi_0)
  &\approx& \bar{\epsilon}(\mu,\xi_0) 
  = \left({m\over 3m_p}\right) \sqrt{4\pi m_p^2 M_\odot\over\sqrt{\alpha}z_t}
   \left(1 - {1\over 12z_t} - {6v_t\over 5z_t}\right) \mu^{1/2} \xi_0^{3/4}
  \nonumber \\
  &\approx& (0.0001815)(1.0235)(0.9985)\, \mu^{0.4951} \xi_0^{0.7418}
  \nonumber \\
  &\approx& (0.0001815)(1.0220)\, \mu^{0.4951} \xi_0^{0.7418},
  \nonumber \\
  &\approx& 0.0001855\, \mu^{0.4951} \xi_0^{0.7418},
 \label{eq:149}
 \end{eqnarray}
where $(1/3)m/m_p$ gives the factor of 0.0001815, the square root (when $z_t$
is evaluated to give 75.770 for $Q_0=M=M_\odot$) gives the factor of 1.0235
that is accidentally close to unity, and the truncated series with the
$O(z_t^{-1})$ corrections gives the factor of 0.9985.  Therefore, the main
cause for making the very conservative upper limit on the efficiency $\epsilon$
much smaller than unity is the small ratio of the mass of the electron to the
mass of the proton, and it turned out coincidentally that for a solar mass
core, the maximum idealized efficiency of converting maximally charged core
energy into outgoing positron energy is within about 2.2\% of one-third this
small mass ratio.

If one multiplies the efficiency $\epsilon$ from Eq. (\ref{eq:147}) by the
mass-energy $M = Mc^2$ of the core, one gets that the total energy emitted in
positrons is
 \begin{equation}
 \epsilon M \approx 3.315\times 10^{50} \left({M\over M_\odot}\right)^{1.495}
  \left({Q_0\over M}\right)^{0.742} \mathrm{ergs},
 \label{eq:150}
 \end{equation}
which for a core of mass $\tilde{M} \equiv m_p^{-2} \approx 1.85327 M_\odot$
and of maximal initial charge ($Q_0=M=\tilde{M}$) is very roughly $m/(3m_p^3) =
\hbar c m/(3Gm_p^3) \sim 6\times 10^{50}$ ergs.  For stellar mass cores, this
is 2-3 orders of magnitude smaller than the energy of gamma-ray bursts
\cite{RSWX,TA}, essentially because the efficiency is 3-4 orders of magnitude
less than unity.

Again, I should emphasize that all of these estimates and calculations give
only very conservative upper limits on the efficiency and energy emitted, since
they all assume that somehow one can hold the charge onto the surface of the
collapsing core even when it is as large as the transition radius $R_t$, which
for $\xi_0 = Q_0/M > 10^{-4}$ would be significantly greater than the size of a
neutron star.  For such a core the electrostatic forces of repulsion on the
excess protons on its surface would be more than 14 orders of magnitude greater
than the gravitational attraction of the core for these protons, and it is hard
to imagine that any forces sufficiently powerful to hold in the protons (such
as nuclear forces at nuclear distances) could be effective in any star larger
than a neutron star.  And if one does take $\xi_0 \sim 10^{-4}$ so that $R_t
\sim 50$ km, somewhat larger than a neutron star size, then the upper limit on
the efficiency of the pair production process would be only of the order of
$2\times 10^{-7}$, which is far too small to give a viable model for gamma ray
bursts.

\section{Annihilation probabilities}

We have taken an idealized model in which originally there is no matter outside
the positively charged collapsing stellar core, but only an electric field
that produces electron-positron pairs there.  We have assumed that the
electrons produced propagate freely to the core surface, and that the positrons
produced propagate freely to radial infinity (both being accelerated by the
electric field, of course).  Now we shall confirm that indeed the annihilation
probabilities are very small for these electrons and positrons, so that indeed
they propagate essentially freely.

The cross section for an electron to annihilate with a positron that has a
large gamma-factor $\gamma \gg 1$ in the frame of the electron is \cite{Dirac}
 \begin{equation}
 \sigma \approx {\pi q^4 \over m^2 \gamma}[\ln{(2\gamma)}-1].
 \label{eq:151}
 \end{equation}
From this, let us calculate the probability $P$ that a positron created very
near the collapsing stellar core surface annihilates with any one of the
incoming electrons as the positron propagates out to radial infinity, as a
function of $v=\sqrt{2M/R}$.  Since the annihilation probability will turn out
to be extremely small, it is sufficient to confirm that by taking the Newtonian
limit (strictly speaking valid only for $v \ll 1$, but here applied to all $v
\leq 1$) for the core collapse and exterior geometry.

Since the electrons and positrons are very rapidly accelerated to near the speed
of light, in the lab frame (the center of mass frame of the collapsing core)
they will have a relative velocity of very nearly $2c=2$, so the annihilation
probability (if small) would be
 \begin{equation}
 P \approx 2\int_R^\infty F \sigma dr
 \label{eq:152}
 \end{equation}
if $F$ were the inward number flux of mono-energetic electrons.

However, the cross section depends on the energies of the outgoing positron and
the incoming electrons, so one needs to divide up the flux of electrons into
partial fluxes according to their energies.  In particular, consider electrons
that are produced between $r_e$ and $r_e + dr_e$ and then propagate inward to
give the partial flux $(dF/dr_e)dr_e$, where for $r_e-R \ll R$
 \begin{eqnarray}
 {dF\over dr_e} &\approx& \mathcal{N}(t,r_e)
 = {m^4\over 4\pi} {e^{-w(t,r_e)}\over w(t,r_e)^2}
 = {q^2Q(t)^2\over4\pi^3 r_e^4}\exp{\left(-{\pi m^2 r_e^2\over qQ(t)}\right)}
 \nonumber \\
 &\approx& {m^4\over 4\pi} {e^{-z}\over z^2}
 \exp{\left[-{2z\over R}(r_e-R)\right]}.
 \label{eq:153}
 \end{eqnarray}
Then the annihilation probability may be written as
 \begin{equation}
 P \approx 2\int_R^\infty dr \int_r^\infty dr_e {dF\over dr_e} \sigma.
 \label{eq:154}
 \end{equation}

At a possible annihilation point $r$ between the core radius $R$ (where the
positron is assumed to be produced, to maximize the annihilation probability
$P$) and the electron production radius $r_e$, the positron and electrons will
have accelerated to the following gamma factors, respectively, in the lab
frame:
 \begin{equation}
 \gamma_+ \approx {qQ\over m}\left({1\over R}-{1\over r}\right)
 \approx {qQ(r-R)\over m R^2} \approx {\pi m\over z}(r-R),
 \label{eq:155}
 \end{equation}
 \begin{equation}
 \gamma_- \approx {qQ\over m}\left({1\over r}-{1\over r_e}\right)
 \approx {qQ(r_e-r)\over m R^2} \approx {\pi m\over z}(r_e-r),
 \label{eq:156}
 \end{equation}
where the final two expressions on the right hand side of each of these
approximate equations are valid for $r_e-R \ll R$, which is where almost all of
the electrons are produced.  For $\gamma_+ \gg 1$ and $\gamma_- \gg 1$, the
relative gamma-factor between the electron and the positron, which goes into
the cross section formula (\ref{eq:151}), is
 \begin{equation}
 \gamma \approx 2\gamma_+\gamma_-.
 \label{eq:157}
 \end{equation}

Now we may see that the exponent in the final exponential of Eq. (\ref{eq:153})
is $-\delta(\gamma_++\gamma_-)$, where
 \begin{equation}
 \delta = {v^2 z^2 \over \pi m M}.
 \label{eq:158}
 \end{equation}
Then in Eq. (\ref{eq:154}) we may also use $r \approx R+(z/\pi m)\gamma_+$, $r_e
\approx r+(z/\pi m)\gamma_-$, $qQ=4\pi m^2 R^2/z$, and $R=2M/v^2$, and
also insert $\gamma$ from Eq. (\ref{eq:157}) into the cross section formula
(\ref{eq:151}) to obtain
 \begin{eqnarray}
 P &\approx& {q^4 e^{-z}\over 4\pi^2}
 \int_1^\infty {d\gamma_+\over \gamma_+} \int_1^\infty {d\gamma_-\over \gamma_-}
 (\ln{\gamma_+} + \ln{\gamma_-} + 2\ln{2} - 1)e^{-\delta(\gamma_++\gamma_-)}
 \nonumber \\
 &\approx& {q^4 e^{-z}\over 4\pi^2}
 \left[\ln^3{\pi m M\over v^2 z^2}+(2\ln{2}-1)\ln^2{\pi mM\over v^2z^2}\right].
 \label{eq:159}
 \end{eqnarray}

When the self-regulation of the charge is effective, the first term on the right
hand side of Eq. (\ref{eq:131}) dominates, so one has
 \begin{equation}
 z^2 e^z \approx {A\mu^2\over v^5} \equiv {q^2 m^2 M^2 \over \pi v^5}. 
 \label{eq:160}
 \end{equation}
Then
 \begin{equation}
 P \approx {q^2 v^5 z^2 \over 4\pi m^2 M^2}
 \left[\ln^3{\pi m M\over v^2 z^2}+(2\ln{2}-1)\ln^2{\pi mM\over v^2z^2}\right].
 \label{eq:161}
 \end{equation}

The maximum probability this formula gives is when $M$ is minimized and $v$ is
maximized, so let us set $M=M_\odot$ as a very conservative minimum value of
the mass of a collapsing core that can form a black hole, and also take $v=1$,
its maximum value outside the black hole.  Then the numerical calculations
described above gave $z \approx 57.6048$, which leads to an annihilation
probability of only
 \begin{equation}
 P \approx 3.23\times 10^{-27},
 \label{eq:162}
 \end{equation}
which is utterly negligible.  Therefore, one may completely neglect the
annihilation of the electrons and positrons produced by the electric field of a
collapsing stellar core.

The main reason that the annihilation probability is so small is that when the
positron and the electron have each accelerated a radial distance in the
electric field that is much greater than the distance $m/(qE) = z/(\pi m)$
needed to give each particle a kinetic energy equal to its rest mass, so that
the positron and electron energies are $E_+ \gg m$ and $E_- \gg m$, the
annihilation cross section given by (\ref{eq:151}),
 \begin{equation}
 \sigma \approx {\pi q^4 \over 2 E_+ E_-}\left(\ln{4E_+E_-\over m^2}-1\right),
 \label{eq:163}
 \end{equation}
drops roughly inversely with each acceleration distance.  Therefore, the
integration over these acceleration distances gives a result only logarithmic
(with the leading term going as the cube of the logarithm) in the effective
total available acceleration distance, which for the region outside a charged
sphere of radius $R$ is roughly $R/(2z)$.  Even for small $z$, so that the
$e^{-z}$ factor in Eq. (\ref{eq:159}) does not give significant damping, one
needs a very large available acceleration distance (or very small $\delta$) in
order for the cube of its logarithm to compensate for the $q^4 = \alpha^2$
factor in Eq. (\ref{eq:159}).

\section{Conclusions}

It does not seem to be at all nearly possible to have astrophysical
dyadospheres (electric fields larger than the critical value for Schwinger pair
production, over macroscopic regions much larger than the regions of high
fields that might conceivably be produced by individual heavy nuclear
collisions).  If, as is most plausible, charge carriers like protons are bound
to an astrophysical object, such as a star or stellar core, primarily by
gravitational forces, then the electric field cannot get within 13 orders of
magnitude of the minimal dyadosphere values.  (First the excess charges will
simply be ejected by the electrostatic repulsion when that exceeds the
gravitational attraction.  Then pair production rates from the macroscopic
electric field, as opposed to that from collisions of individual particles,
will be trillions of orders of magnitude below dyadosphere values and so will
be completely negligible.)

Even in the implausible scenario in which the excess charge carriers are bound
by nuclear forces to a collapsing stellar core, I have shown here in a simple
spherically symmetric model that the electric field has a very conservative
maximum value that is more than a factor of 18 below the minimal dyadosphere
value.  Because the pair production rate is essentially exponential in the
negative inverse of the electric field, the upper limit of the pair production
rate in even this implausible scenario is more than 26 orders of magnitude
below the minimal dyadosphere values.

The idealized implausible scenario considered here, to give this very
conservative upper bound on the pair production rate, had the maximal amount of
charge somehow bound to the surface of an idealized stellar core with maximal
initial charge that undergoes free fall collapse from radial infinity in the
Schwarzschild metric (conservatively ignoring the fact that an actual
astrophysical collapse would start at finite radius and fall in slower, giving
more time for discharge, and the fact that if the initial charge were maximal,
the strong electric field would modify the geometry and also give electrostatic
repulsion of the core surface, both of which would also slow down the collapse
and lead to greater discharge and smaller electric fields).  This scenario led
to the maximal electric field (occurring when the freely collapsing core enters
the event horizon of the black hole that would form) being less than 5.5\% of
dyadosphere values.  Although the pair production rate is always less than
$10^{-26}$ that of dyadosphere values, in this implausible scenario of having
no other mechanism of discharging the core, there would be enough pair
production to keep the electric fields always more than 18 times smaller than
dyadosphere values.

Although astrophysical dyadospheres do not form, in this idealized implausible
scenario there is significant pair production (though at astrophysical rates
that are much, much lower than dyadosphere rates), and it may be of interest to
calculate what fraction of the total mass-energy $M$ of the collapsing stellar
core would be converted into pairs.  Here it was found that this efficiency,
even under the highly idealized conditions of having maximal initial charge at
such large radii that it seems inconceivable that the charge carriers could be
sufficiently bound to such objects so much larger than neutron stars, was
always much less than unity for collapsing objects with much less mass than
three million solar masses:  the efficiency was very conservatively bounded by
$2\times 10^{-4}\sqrt{M/M_\odot}$.  Therefore, even these idealized charged
collapsing objects, unless they were enormously more massive than the sun,
would not produce enough energy in outgoing charged particles to be consistent
with the observed gamma ray bursts.

It would of course be of interest to calculate the upper limits on the electric
field and on the maximum pair production rate for models in which one relaxed
the spherical symmetry.  Although I would readily admit that I do not have a
rigorous proof that the pair production rate cannot be higher then, the general
arguments of Subsection 2.3 strongly suggest that it would be very surprising
if it could be larger by more than about one order of magnitude.  Therefore,
the spherically symmetric model analyzed here compel me to conjecture that the
maximal pair production rates achievable by macroscopic astrophysical electric
fields are more than 25 orders of magnitude below that of hypothetical
dyadosphere values.

In conclusion, dyadospheres almost certainly cannot form astrophysically, and
the much weaker pair production rates that might occur, under highly idealized
and implausible scenarios, do not seem sufficient for giving viable models of
gamma ray bursts.  For a shorter version of this work for a conference
proceedings, see \cite{dy}.

%\newpage
\section*{Acknowledgments}

I am grateful for the Asia Pacific Center of Theoretical Physics and for Kunsan
University for enabling me to participate in the 9th Italian-Korean Symposium
on Relativistic Astrophysics, 2005 July 19-24, Seoul, South Korea, and Mt.
Kumgang, North Korea, where I learned about dyadospheres.  I am also thankful
for the hospitality of Beijing Normal University, People's Republic of China,
where my preliminary literature search and calculations were performed.  I
appreciated being able to discuss my work in progress at the VII Asia-Pacific
International Conference on Gravitation and Astrophysics (ICGA7), 2005 November
23-26, National Central University, Jhongli, Taiwan, Republic of China.  This
work was also supported in part by the Natural Sciences and Engineering
Research Council of Canada.

\end{document}